\documentclass[prd,aps,showpacs,nofootinbib]{revtex4}

\usepackage{amssymb,graphicx}
\usepackage{epsfig}
\newcommand{\re}{\mbox{Re}}
\newcommand{\im}{\mbox{Im}}
\newcommand{\gz}{\mbox{\em \r{g}\hspace{0.3mm}}}  % Zero order approximation.
\newcommand{\Rz}{\mbox{\em \r{R}}}

\newcommand{\nz}{\mbox{\em \r{n}\hspace{0.3mm}}}
\newcommand{\sz}{\mbox{\em \r{s}\hspace{0.3mm}}}
\newcommand{\Dz}{\mbox{\em \r{D}\hspace{0.3mm}}}
\newcommand{\nablaz}{\nabla\hspace{-0.27cm}{}^{\mbox{\r{~}}}{}\hspace{-0.22cm}}

\newcommand{\Gammaz}{\Gamma\hspace{-0.25cm}{}^{\mbox{\r{~}}}{}\hspace{-0.12cm}}
\newcommand{\CKz}{{\cal K}\hspace{-0.25cm}{}^{\mbox{\r{~}}}{}\hspace{-0.12cm}}

\newcommand{\Natural}{\mathbb N}
\newcommand{\Real}{\mathbb{R}}
\newcommand{\Complex}{\mathbb{C}}

\newcommand{\hateq}{\; \hat{=}\; }
\newcommand{\half}{\textstyle \frac{1}{2}}
\newcommand{\Or}{\mathcal{O}}

\newcommand{\plot}[1]{     
  \begin{minipage}[t]{0.49\textwidth}
    \includegraphics[scale=0.32]{#1}
    \medskip
  \end{minipage}
}

\begin{document}

%%%%%%%%%%%%%%%%%
%%%   TITLE   %%%
%%%%%%%%%%%%%%%%%

\title{Outer boundary conditions for Einstein's field equations in
harmonic coordinates}

\author{Milton Ruiz$^{1}$, Oliver Rinne$^{2,4}$ and Olivier Sarbach$^{3}$}

\affiliation{$^{1}$Instituto de Ciencias Nucleares, Universidad Nacional
Aut\'onoma de M\'exico, A.P. 70-543, M\'exico D.F. 04510, M\'exico}

\affiliation{$^2$Theoretical Astrophysics 130-33,
California Institute of Technology,
1200 East California Boulevard,
Pasadena, California 91125--0001, USA}

\affiliation{$^3$Instituto de F\a'{\i}sica y Matem\a'aticas,
Universidad Michoacana de San Nicol\a'as de Hidalgo, Edificio C-3,
Cd. Universitaria. C. P. 58040 Morelia, Michoac\a'an, M\'exico}

\affiliation{$^4$Current address: DAMTP, CMS, Wilberforce Road,
  Cambridge CB3 0WA, UK and King's College, Cambridge CB2 1ST, UK}

%%%%%%%%%%%%%%%%
%%%   DATE   %%%
%%%%%%%%%%%%%%%%

\date{\today}

%%%%%%%%%%%%%%%%%%%%%%%%%%%%%%%%
\begin{abstract}
%%%%%%%%%%%%%%%%%%%%%%%%%%%%%%%%
We analyze Einstein's vacuum field equations in generalized harmonic
coordinates on a compact spatial domain with boundaries. We specify a
class of boundary conditions which is constraint-preserving and
sufficiently general to include recent proposals for reducing the
amount of spurious reflections of gravitational radiation. In
particular, our class comprises the boundary conditions recently
proposed by Kreiss and Winicour, a geometric modification thereof, the
freezing-$\Psi_0$ boundary condition and the hierarchy of absorbing
boundary conditions introduced by Buchman and Sarbach. Using the
recent technique developed by Kreiss and Winicour based on an
appropriate reduction to a pseudo-differential first order system, we
prove well posedness of the resulting initial-boundary value problem
in the frozen coefficient approximation. In view of the theory of
pseudo-differential operators it is expected that the full nonlinear
problem is also well posed. Furthermore, we implement some of our
boundary conditions numerically and study their effectiveness in a
test problem consisting of a perturbed Schwarzschild black hole.
\end{abstract}

% 04.20.-q: Classical GR
% 04.20.Ex: Initial value problem, existence and uniqueness of solutions
% 04.25.-g: Approximation methods; equations of motion
% 04.25.Dm: Numerical relativity
% 04.40.-b: Self-gravitating systems, continuous media and 
%           classical fields in curved spacetime
% 04.70.-s: Physics of black holes
% 97.60.Lf: Astronomy: Late stage of star evolution: black holes

\maketitle

%%%%%%%%%%%%%%%%%%%%%%%%%%%%%%%%%%%%%%%%%%%%%%%%%%%%%%%%%%%%%%
\section{Introduction}
\label{Sect:Intro}
%%%%%%%%%%%%%%%%%%%%%%%%%%%%%%%%%%%%%%%%%%%%%%%%%%%%%%%%%%%%%%

A common way to deal with the numerical simulation of wave propagation
on an infinite domain is to replace the latter by a finite
computational domain $\Sigma$ with artificial boundary
$\partial\Sigma$. Boundary conditions at $\partial\Sigma$ must then be
specified such that the resulting Cauchy problem is well
posed. Additionally, the artificial boundary must be as transparent as
possible to the physical problem on the infinite domain in the sense
that it does not introduce too much spurious reflection from the
boundary surface. The construction of such absorbing boundary
conditions has received significant attention for wave problems in
acoustics, electromagnetism, meteorology, and solid geophysics (see
\cite{Givoli91} for a review).

In this article, we construct absorbing boundary conditions for
Einstein's field equations in generalized harmonic coordinates. In
these coordinates, one obtains a system of ten coupled quasi-linear
wave equations for the ten components of the metric field. Therefore,
ten boundary conditions must be imposed. However, it is important to
realize that not all ten components of the metric represent physical
degrees of freedom, so that one cannot simply apply the known results
from the scalar wave equation directly to the ten wave equations for
the metric components. Instead, one has to take into account the fact
that the metric is subject to the harmonic condition which yields four
constraints. In the absence of boundaries, it is possible to show that
it is sufficient to solve these constraints along with their time
derivatives on an initial Cauchy surface; the Bianchi identities and
the evolution equations then guarantee that the constraints are
satisfied everywhere and at each time. When timelike boundaries are
present, four constraint-preserving boundary conditions need to be
specified at $\partial\Sigma$ in order to insure that no
constraint-violating modes propagate into the computational
domain. This reduces the ten degrees of freedom to six.  Another four
degrees of freedom are related to the residual gauge freedom in
choosing harmonic coordinates and fixing the geometry of the boundary
surface. Therefore, one is left with two degrees of freedom which are
related to the gravitational radiation. The challenge, then, is to
specify boundary conditions at $\partial\Sigma$ which preserve the
constraints, form a well posed initial-boundary value problem (IBVP)
and minimize spurious reflections of gravitational radiation from the
boundary.  Additionally, one might want to require that gauge and
constraint-violating modes propagate out of the domain without too
much reflection.

Constraint-preserving boundary conditions for the harmonic system have
been proposed before in \cite{Szilagyi:2002kv, Szilagyi:2001fy,
Lindblom06, Kreiss:2006mi} and tested numerically in
\cite{Szilagyi:2001fy, Babiuc:2006wk, Motamed:2006uw, Babiuc:2006ik,
Rinne:2006vv, Rinne:2007ui}. The boundary conditions of
\cite{Szilagyi:2002kv, Szilagyi:2001fy} are a combination of
homogeneous Dirichlet and Neumann conditions, for which well posedness
can be shown by standard techniques. On the other hand, the conditions
of \cite{Szilagyi:2002kv, Szilagyi:2001fy} are likely to yield large
spurious reflections of gravitational radiation and probably do not
give a good approximation to the solution on the unbounded
domain. Inhomogeneous boundary conditions which allow for a matching
to a characteristic code are also considered in
\cite{Szilagyi:2001fy}, but in this case the well posedness of the
problem is not established. The boundary conditions of
\cite{Kreiss:2006mi} are of the Sommerfeld type, and the well
posedness of the resulting IBVP has been shown, at least in the frozen
coefficient approximation. Finally, the boundary conditions presented
in \cite{Lindblom06} contain second derivatives of the metric fields
and freeze the Weyl scalar $\Psi_0$ to its initial value. As discussed
below, this condition serves as a good first approximation to an
absorbing condition. In this paper, we generalize the first order
boundary conditions of \cite{Kreiss:2006mi} to the full nonlinear
case, and also obtain more general second order boundary conditions
that are very similar to the conditions presented in
\cite{Lindblom06}. Furthermore, we obtain a class of
constraint-preserving higher order boundary conditions which are
flexible enough to incorporate the recently proposed hierarchy of
absorbing outer boundary conditions proposed in \cite{Buchman:2006xf,
Buchman:2007pj}.

There has been a considerable amount of work on constructing well
posed constraint-preserving boundary conditions for Einstein's field
equations. A well posed IBVP for Einstein's vacuum equations was
presented in Ref. \cite{Friedrich99}. This work, which is based on a
tetrad formulation, recasts the evolution equations into first order
symmetric hyperbolic quasilinear form with maximally dissipative
boundary conditions \cite{Friedrichs58,Lax60}, for which (local in
time) well posedness is guaranteed \cite{Secchi96}. There has been a
substantial effort to obtain well posed formulations for the more
commonly used metric formulations of gravity using similar
mathematical techniques (see
\cite{Szilagyi:2001fy,Calabrese:2002xy,Gundlach:2004jp,Nagy:2006pr}
for partial results). A different technique for showing the well
posedness of the IBVP is based on the frozen coefficient principle
where one freezes the coefficients of the evolution and boundary
operators. In this way, the problem is simplified to a linear,
constant coefficient problem on the half-space which can be solved
explicitly by using a Fourier-Laplace transformation
\cite{Kreiss89}. This method yields a simple algebraic condition (the
determinant condition) which is necessary for the well posedness of
the IBVP. Sufficient conditions for the well posedness of the frozen
coefficient problem were developed by Kreiss \cite{Kreiss70}. Kreiss'
theorem provides a stronger form of the determinant condition whose
satisfaction leads to well posedness if the evolution system is
strictly hyperbolic. One of the key results in \cite{Kreiss70} is the
construction of a smooth symmetrizer for the problem for which well
posedness can be shown via an energy estimate in the frequency
domain. Using the theory of pseudo-differential operators, it is
expected that the verification of Kreiss' condition also leads to well
posedness for quasilinear problems, such as Einstein's field
equations. Work based on the verification of the Kreiss condition in
the Einstein case is given in \cite{Stewart98,Rinne:2006vv}. In
particular, generalized harmonic gauge and second order boundary
conditions similar to the ones considered here were analyzed in
\cite{Rinne:2006vv}. However, since in those cases the evolution
system is not strictly hyperbolic, it is not clear if those results
are sufficient for well posedness. Kreiss' theorem was generalized to
symmetric hyperbolic systems in \cite{Majda75} but their treatment
assumed maximally dissipative boundary conditions. On the other hand,
the recent work by Kreiss and Winicour \cite{Kreiss:2006mi} introduces
a new pseudo-differential first order reduction of the wave equation
which leads to a strictly hyperbolic system. Using this reduction they
are able to verify Kreiss' condition and in this way show well
posedness of the IBVP for Einstein's field equations in harmonic
coordinates.

We use this frozen coefficient technique in order to analyze the well
posedness of the IBVPs resulting from our different boundary
conditions. To this end, we consider small amplitude, high frequency
perturbations of a given smooth background solution. In this case, the
problem reduces to a system of ten decoupled wave equations on a
frozen metric background on the half space with linear boundary
conditions. By performing a suitable coordinate transformation which
leaves the half space domain invariant, one can obtain all the metric
coefficients to be those of the flat metric {\em with the exception of
the component of the shift normal to the boundary} (see also
\cite{Rinne:2006vv}). We then prove using the Fourier-Laplace
technique and Kreiss' theorem that our frozen coefficient problem is
well posed. In view of the existence of a smooth symmetrizer and the
theory of pseudo-differential operators \cite{Taylor99b}, it is
expected that one can show well posedness of the full nonlinear
problem as well.

This paper is organized as follows. In Sec.~\ref{Sect:FieldEqs} we
summarize the generalized harmonic formulation of general relativity.
In Sec.~\ref{Sect:OuterBC} we study the resulting wave evolution
equations for the ten components of the metric on a manifold of the
form $M = [0,T] \times \Sigma$, where $\Sigma$ is a three-dimensional
compact manifold with smooth boundary $\partial\Sigma$, and we present
several possibilities for first, second and higher order boundary
conditions, where here the order refers to the highest derivative of
the metric appearing in the condition. In the first order case, in
Sec.~\ref{Sect:OuterBC_first}, we use the harmonic constraint to
impose four Dirichlet boundary conditions for the constraint
propagation system. Next, we specify two boundary conditions in terms
of the shear of the outgoing null congruence associated with the
two-dimensional cross sections of the boundary surface. These
conditions are a geometric modification of the boundary conditions
presented by Kreiss and Winicour \cite{Kreiss:2006mi}. Finally, we
specify four more boundary conditions with some absorbing properties
on the gauge modes corresponding to the residual gauge freedom.  In
Sec.~\ref{Sect:OuterBC_second} we present second order boundary
conditions. One of the advantages of allowing for second derivatives
of the metric is that one can formulate boundary conditions for the
Weyl curvature scalar $\Psi_0$, which has some attractive properties.
First, $\Psi_0$ (with respect to a suitably chosen tetrad) represents
the incoming radiation at past null infinity. Second, the Weyl tensor
from which $\Psi_0$ is constructed is a gauge-invariant quantity in
the weak field limit of gravity. Third, if spacetime is a small
perturbation of a Schwarzschild black hole, as is the case near the
boundary if the boundary is far enough from the strong field region,
$\Psi_0$ (with respect to a tetrad adapted to the Schwarzschild
background) is invariant with respect to infinitesimal coordinate
transformations and tetrad rotations. A boundary condition considered
in the literature is the so-called freezing-$\Psi_0$ condition
\cite{Friedrich99,Bardeen:2001xx,Sarbach:2004rv,Lindblom05,Lindblom06,Rinne:2006vv,
Nagy:2006pr,Rinne:2007ui}, which freezes $\Psi_0$ to its initial
value. An estimate for the amount of spurious reflections of
gravitational radiation was given in \cite{Buchman:2006xf}. There, a
new hierarchy ${\cal B}_L$, $L=1,2,3,...$, of conditions on $\Psi_0$
was also derived with the property of being perfectly absorbing for
linearized gravitational waves with angular momentum number smaller
than or equal to $L$. Generalizations of these conditions which take
into account correction terms from the curvature were presented in
\cite{Buchman:2007pj}. In order to incorporate these conditions into
our analysis, we consider boundary conditions of arbitrarily high
order in Sec.~\ref{Sect:OuterBC_third}. In Sec.~\ref{Sect:WellPosedness} we
use the method by Kreiss and Winicour to show the well posedness of
the resulting IBVPs in the frozen coefficient approximation. In
particular, we allow for a non-trivial shift vector, which is
important in view of the generalization to the quasi-linear case. In
this sense, our results generalize the work in
Ref. \cite{Kreiss:2006mi} to non-trivial shifts and boundary
conditions of arbitrarily high order. Next, in
Sec.~\ref{Sect:Quality}, we obtain estimates for the amount of
spurious reflections for the boundary conditions constructed in this
article and perform numerical tests based on a perturbed Schwarzschild
black hole. Finally, we conclude in Sec.~\ref{Sect:Conclusions}.

%%%%%%%%%%%%%%%%%%%%%%%%%%%%%%%%%%%%%%%%%%%%%%%%%%%%%%%%%%%%%%
\section{The field equations in generalized harmonic coordinates}
\label{Sect:FieldEqs}
%%%%%%%%%%%%%%%%%%%%%%%%%%%%%%%%%%%%%%%%%%%%%%%%%%%%%%%%%%%%%%

In this section we review the formulation of Einstein's field
equations in generalized harmonic coordinates
\cite{Friedrich85,Friedrich96}
\begin{equation}
H^c = \Box_g x^c = g^{ab}\Gamma^c{}_{ab}\; ,
\label{Eq:HarmonicCoords}
\end{equation}
where $H^c$ are given functions on $M$, $g_{ab}$ is the spacetime
metric and $\Box_g = -g^{ab}\nabla_a\nabla_b$ and $\Gamma^c{}_{ab}$
denote the corresponding d'Alembertian operator and Christoffel
symbols, respectively. Instead of adopting the gauge condition
(\ref{Eq:HarmonicCoords}), we find it convenient to choose a {\em
fixed} background manifold $(M,\gz_{ab})$ and replace
(\ref{Eq:HarmonicCoords}) by ${\cal C}^c = 0$, where the vector field
${\cal C}^c$ is given by
\begin{equation}
{\cal C}^c = g^{ab}\left( \Gamma^c{}_{ab} - \Gammaz^c{}_{ab} \right) - H^c.
\label{Eq:HarmConstr}
\end{equation}
Here, $H^c$ is a given vector field on $M$ and $\Gammaz^c{}_{ab}$ are
the Christoffel symbols corresponding to the background metric
$\gz_{ab}$. In the particular case where the background manifold is
Minkowski spacetime and standard Cartesian coordinates are chosen on
$M$, $\Gammaz^c{}_{ab}$ vanishes, and the condition ${\cal C}^c = 0$
reduces to Eq. (\ref{Eq:HarmonicCoords}). The advantage of using
${\cal C}^c$ is that, unlike $\Box_g x^c$, it transforms as a vector
field since the difference between the two Christoffel symbols,
\begin{equation}
C^c{}_{ab} \equiv \Gamma^c{}_{ab} - \Gammaz^c{}_{ab}
 = \frac{1}{2} g^{cd}\left( \nablaz_{a} h_{bd} + \nablaz_{b} h_{ad} 
 - \nablaz_d h_{ab} \right),
\label{Eq:Christoffel}
\end{equation}
forms a tensor field. Here and in the following, $h_{ab} = g_{ab} -
\gz_{ab}$ denotes the difference between the dynamical metric $g_{ab}$
and the background metric $\gz_{ab}$. Since $\nablaz_c\gz_{ab} = 0$,
one could also replace $\nablaz_c h_{ab}$ with $\nablaz_c g_{ab}$ in
Eq. (\ref{Eq:Christoffel}); however, we prefer to express our
equations in terms of the difference field $h_{ab}$ instead of the
metric $g_{ab}$. A condition that is related to ${\cal C}^c = 0$ was
used in \cite{Andersson03} for imposing spatial harmonic coordinates.

The curvature tensor corresponding to the metric $g_{ab}$ can be written
as
\begin{equation}
R^a{}_{bcd} = \Rz^a{}_{bcd} + 2\nablaz_{[c} C^a{}_{d]b} 
            + 2 C^a{}_{e[c} C^e{}_{d]b}\; ,
\label{Eq:Curvature}
\end{equation}
where $\Rz^a{}_{bcd}$ denotes the curvature tensor with respect to the
background metric $\gz_{ab}$. Inserting Eq. (\ref{Eq:Christoffel})
into (\ref{Eq:Curvature}) and using $\nablaz_a g^{cd} = -C^c{}_{ab}
g^{bd} - C^d{}_{ab} g^{bc}$, one obtains,
\begin{eqnarray}
R_{abcd} &=& \frac{1}{2} \left( 
   \nablaz_c\nablaz_b h_{ad} - \nablaz_d\nablaz_b h_{ac} 
 + \nablaz_d\nablaz_a h_{cb} - \nablaz_c\nablaz_a h_{bd} \right)
\nonumber\\
 &+& g_{ef} C^e{}_{bc} C^f{}_{ad} - g_{ef} C^e{}_{bd} C^f{}_{ac} 
  + \frac{1}{2}\left( g_{ae}\Rz^e{}_{bcd} - g_{be}\Rz^e{}_{acd} \right).
\label{Eq:Riemann}
\end{eqnarray}
From this, we obtain the corresponding expression for the Ricci
tensor,
\begin{eqnarray}
R_{ab} &=& \frac{1}{2} g^{cd}\left( 
 -\nablaz_c\nablaz_d h_{ab} - \nablaz_a\nablaz_b h_{cd} 
 + \nablaz_a\nablaz_c h_{bd} + \nablaz_b\nablaz_c h_{ad} \right)
\nonumber\\
 &+& g_{ef} g^{cd}\left( C^e{}_{ac} C^f{}_{bd} - C^e{}_{ab} C^f{}_{cd} \right)
-  g^{cd} \Rz^e{}_{cd(a} g_{b)e} \; .
\label{Eq:Ricci}
\end{eqnarray}
On the other hand, we have
\begin{eqnarray}
\nabla_a {\cal C}_b &=& g^{cd}\left( 
  \nablaz_a\nablaz_c h_{bd} - \frac{1}{2}\, \nablaz_a\nablaz_b h_{cd} 
\right)
 - 2\,C^c{}_{da}\, g_{be}\, C^e{}_{cf}\,g^{df}
\nonumber\\
&-& g_{ef}\, g^{cd}\,C^e{}_{ab}\, C^f{}_{cd}
 - \nabla_a H_b\; .
\label{Eq:GradC}
\end{eqnarray}
Subtracting the symmetric part of Eq. (\ref{Eq:GradC}) from
Eq. (\ref{Eq:Ricci}) we obtain
\begin{eqnarray}
{\cal E}_{ab} &=& g^{cd}\nablaz_c\nablaz_d h_{ab}
 - 2\, g_{ef} g^{cd} C^e{}_{ac} C^f{}_{bd} 
 - 4\, C^c{}_{d(a} g_{b)e} C^e{}_{cf} g^{df} 
\nonumber\\
 &+& 2\, g^{cd} \Rz^e{}_{cd(a} g_{b)e}
 - 2\,\nabla_{(a} H_{b)}\,,
\label{Eq:DefE}
\end{eqnarray}
where ${\cal E}_{ab} \equiv -2R_{ab} + 2\nabla_{(a} {\cal
C}_{b)}$. Using the twice contracted Bianchi identities we also obtain,
\begin{equation}
\nabla^b\left( {\cal E}_{ab} - \frac{1}{2} g_{ab} g^{cd} {\cal E}_{cd} \right)
 = \nabla^b\nabla_b {\cal C}_a + R_a{}^b {\cal C}_b\; .
\label{Eq:Bianchi}
\end{equation}

The Cauchy problem for Einstein's vacuum equations in generalized
harmonic coordinates on an infinite domain of the form $M = [0,T]
\times \Real^3$ can be formulated in the following two steps. First,
specify initial data on the hypersurface $\Sigma_0 := \{ 0 \} \times
\Real^3$. For this, let $n^a$ and $\nz^a$ denote the future-pointing
unit normals to $\Sigma_0$ with respect to the metrics $g_{ab}$ and
$\gz_{ab}$, respectively. Notice that the corresponding one-forms,
$n_a$ and $\nz_a$, are proportional to each other: $n_a = \alpha\,
\nz_a$. Decompose the dynamical metric in the form
\begin{displaymath}
g_{ab} = -\alpha^2 \nz_a\nz_b 
 + \gamma_{cd}\left( \delta^c{}_a - \beta^c\nz_a \right)
              \left( \delta^d{}_b - \beta^d\nz_b \right),
\end{displaymath}
where $\gamma_{cd}\nz^c = 0$, $\beta^c\nz_c = 0$. The pull-back of
$\gamma_{ab}$ on $\Sigma_0$ is the metric $\bar{g}_{ab}$ induced by
$g_{ab}$ on $\Sigma_0$, and $\alpha$ and $\beta^a$ are generalized
lapse and shift. Next, solve the Hamiltonian and momentum constraints
$n^a(2R_{ab} - g_{ab} g^{cd} R_{cd}) = 0$ for the induced metric
$\bar{g}_{ab}$ and the extrinsic curvature, $k_{ab}$. Then, solve the
equation ${\cal C}^a = 0$ which yields \cite{Pretorius:2004jg,Lindblom06}
\begin{eqnarray}
\dot{\alpha} &=& \beta^a\Dz_a \alpha -\alpha^2\gamma^{ab} k_{ab} 
 - \alpha(\nz_a H^a + \gamma_{ab} \beta^a H^b),
\label{Eq:HarmConstrAlphaDot}\\
\dot{\beta}^a &=& \beta^b\Dz_b\beta^a - \alpha\gamma^{ab}\Dz_b\alpha
 + \alpha^2\gamma^{ab}\gamma^{cd} \left[
    \Dz_c\gamma_{bd} - \frac{1}{2}\Dz_b\gamma_{cd} 
 - \gamma_{bc} \gamma_{de} H^e \right],
\label{Eq:HarmConstrBetaDot}
\end{eqnarray}
where a dot refers to the Lie derivative with respect to $\nz^a$, and
$\Dz$ denotes the connection on $\Sigma_0$ which is induced by
$\nablaz$. Here, we have assumed that $\nablaz_a\nz_b = 0$ for
simplicity. Equations
(\ref{Eq:HarmConstrAlphaDot},\ref{Eq:HarmConstrBetaDot}) can be solved
by either first choosing $\alpha$, $\beta^a$ and $H^a$ which fixes
$\dot{\alpha}$ and $\dot{\beta}^a$ or by choosing $\alpha$, $\beta$,
$\dot{\alpha}$ and $\dot{\beta}^a$ and solving for the vector field
$H^a$. The quantities $\alpha$, $\dot{\alpha}$, $\beta^a$,
$\dot{\beta}^a$, $\gamma_{ab}$ and $k_{ab}$ determine the initial data
for $h_{ab}$ and $\dot{h}_{ab} \equiv \pounds_{\nz} h_{ab}$ by taking
into account that
\begin{displaymath}
\dot{\gamma}_{ab} = \left( -2\alpha k_{cd} + \pounds_\beta\gamma_{cd} \right)
(\delta^c{}_a + \nz^c\nz_a)(\delta^d{}_b + \nz^d\nz_b).
\end{displaymath}

The second step consists in finding a solution $h_{ab}$ on $M$ of the
nonlinear wave equation (\ref{Eq:DefE}) with ${\cal E}_{ab} = 0$
subject to the initial data specified on $\Sigma_0$. The results in
\cite{Foures-Bruhat52,Fischer72} show that there exists a unique
solution to this problem, at least if $T$ is small enough. Finally, we
observe that the evolution equations (\ref{Eq:DefE}) with ${\cal
E}_{ab} = 0$ imply that \cite{Lindblom06}
\begin{equation}
n^a\nabla_a {\cal C}^b = n_a(2R^{ab} - g^{ab} g_{cd} R^{cd}) 
 + (\gamma^{ac} n^b - n^a\gamma^{bc})\nabla_c {\cal C}_a\; .
\end{equation}
Since on $\Sigma_0$ the Hamiltonian and momentum constraints and
${\cal C}^a = 0$ are satisfied, it follows that $\dot{\cal C}^a = 0$.
The evolution system for the harmonic constraint variables,
Eq. (\ref{Eq:Bianchi}) with ${\cal E}_{ab} = 0$, then guarantees that
${\cal C}_a = 0$ everywhere on $M$ since it has the form of a linear
homogeneous wave equation for ${\cal C}_a$ with trivial initial data
${\cal C}_a = 0$, $\dot{\cal C}_a = 0$. In the next section, we
analyze the Cauchy problem when the infinite domain $\Real^3$ is
replaced by a bounded domain $\Sigma$.

%%%%%%%%%%%%%%%%%%%%%%%%%%%%%%%%%%%%%%%%%%%%%%%%%%%%%%%%%%%%%%
\section{Outer boundary conditions}
\label{Sect:OuterBC}
%%%%%%%%%%%%%%%%%%%%%%%%%%%%%%%%%%%%%%%%%%%%%%%%%%%%%%%%%%%%%%

We wish to study the evolution equations (\ref{Eq:DefE}) with ${\cal
E}_{ab} = 0$ on a manifold of the form $M = [0,T] \times \Sigma$,
where $\Sigma$ is a three-dimensional compact manifold with smooth
boundary $\partial\Sigma$. We assume that the boundary surface ${\cal
T} = [0,T] \times \partial\Sigma$ is timelike and that the
three-dimensional surfaces $\Sigma_t = \{ t \} \times \Sigma$ are
spacelike. The cross section $S_t = \{ t \} \times \partial\Sigma$
constitutes the boundary of $\Sigma_t$. For the following, let $n^a$
denote the future-pointing unit normal to the time-slices $\Sigma_t$
and $s^a$ the unit outward normal to the two surface $S_t$ as embedded
in $\Sigma_t$. These vector fields are defined with respect to the
dynamical metric $g_{ab}$ and {\em not} the background metric
$\gz_{ab}$. Therefore,
\begin{displaymath}
g_{ab} n^a n^b = -1\,, \qquad
g_{ab} n^a s^b = 0\,, \qquad
g_{ab} s^a s^b = 1\,.
\end{displaymath}
The vector fields $n^a$ and $s^a$ allow us to construct a
Newman-Penrose null tetrad $\{ l^a, k^a, m^a ,\bar{m}^a \}$ 
\begin{eqnarray}
l^a &=& \frac{1}{\sqrt{2}}\left( n^a + s^a \right), \qquad
k^a = \frac{1}{\sqrt{2}}\left( n^a - s^a \right), \qquad
\label{Eq:NPTetrad1}\\
m^a &=& \frac{1}{\sqrt{2}}\left( v^a + i\, w^a \right), \qquad
\bar{m}^a = \frac{1}{\sqrt{2}}\left( v^a - i\, w^a \right),
\label{Eq:NPTetrad2}
\end{eqnarray}
where $v^a$ and $w^a$ are two mutually orthogonal unit vector fields
which are normal to $n^a$ and $s^a$ (with respect to the metric
$g_{ab}$). Notice that this null tetrad is naturally adapted to the
two-surfaces $S_t$; it is unique up to rescaling of the real null
vectors $l^a $ and $k^a$ and up to a rotation $m^a \mapsto
e^{i\varphi} m^a$, $\bar{m}^a \mapsto e^{-i\varphi}\bar{m}^a$ of the
complex null vectors $m^a$ and $\bar{m}^a$ about an angle $\varphi$.

Since the evolution equations have the form of ten wave equations (see
Eq. (\ref{Eq:DefE})), we need to specify ten boundary conditions on
${\cal T}$. These ten boundary conditions can be divided into
constraint-preserving boundary conditions, boundary conditions
controlling the physical radiation, and boundary conditions that
control the gauge freedom. Constraint-preserving boundary conditions
make sure that solutions with constraint-satisfying initial data
satisfy the constraints for each $0 < t < T$. Since there are four
constraints, namely ${\cal C}^c = 0$, and these constraints obey a set
of wave equations on their own (see Eq. (\ref{Eq:Bianchi})), there are
four constraint-preserving boundary conditions. Gravitational
radiation has two degrees of freedom, so we need to provide two
boundary conditions responsible for controlling the physical
radiation. The remaining four boundary conditions control the gauge
freedom.

In the following, we discuss several possibilities for fixing such
boundary conditions. We divide them into first, second and higher
order boundary conditions, where here the order refers to the highest
number of derivatives of $h_{ab}$ appearing in the boundary
conditions. These families of boundary conditions are discussed
next. In Sec.~\ref{Sect:WellPosedness}, the well posedness of
the resulting IBVPs in the frozen coefficient approximation is proven.

%%%%%%%%%%%%%%%%%%%%%%%%%%%%%%%%%%%%%%%%%%%%%%%%%%%%%%%%%%%%%%
\subsection{First order boundary conditions}
\label{Sect:OuterBC_first}
%%%%%%%%%%%%%%%%%%%%%%%%%%%%%%%%%%%%%%%%%%%%%%%%%%%%%%%%%%%%%%

In the first order case, constraint-preserving boundary conditions are
specified through
\begin{equation}
{\cal C}_c \equiv 
 g^{ab}\left( \nablaz_a h_{bc} - \frac{1}{2}\nablaz_c h_{ab} \right) - H_c 
 \hateq 0,
\label{Eq:FirstOrderCPBC}
\end{equation}
where here and in the following, the notation $\hateq$ means an
equality which holds on ${\cal T}$ only. These four boundary
conditions are Dirichlet conditions for the constraint propagation
system, Eq. (\ref{Eq:Bianchi}) with ${\cal E}_{ab} = 0$. Since this
system is a linear homogeneous wave equation for ${\cal C}_a$ on the
curved background $(M,g_{ab})$, these boundary conditions imply (by
uniqueness) that solutions of this system with trivial initial data
${\cal C}_a = 0$, $\dot{\cal C}_a = 0$ are identically zero. Using
$g^{ab} = -2\,l^{(a} k^{b)} + 2\,m^{(a} \bar{m}^{b)}$, the four
conditions (\ref{Eq:FirstOrderCPBC}) are equivalent to
\begin{eqnarray}
0 &\hateq &{\cal C}_c l^c \hateq
 - D_{lm\bar{m}} - D_{kll} + D_{m\bar{m}l} 
+ D_{\bar{m}ml} - H_l\; ,
\label{Eq:FirstOrderBC1}\\
0 &\hateq& {\cal C}_c k^c \hateq 
 - D_{lkk} - D_{km\bar{m}} + D_{m\bar{m}k} 
+ D_{\bar{m}mk} - H_k\; ,
\label{Eq:FirstOrderBC2}\\
0& \hateq &{\cal C}_c m^c \hateq
 - D_{lkm} - D_{klm} + D_{mlk} 
+ D_{\bar{m}mm} - H_m\; ,
\label{Eq:FirstOrderBC3}
\end{eqnarray}
where we have defined the tensor field $D_{cab} \equiv \nablaz_c
h_{ab}$ and where the indices $l$, $k$, $m$, $\bar{m}$ refer to
contraction with $l^a$, $k^a$, $m^a$ and $\bar{m}^a$,
respectively. Notice that equation (\ref{Eq:FirstOrderBC3}) comprises
two real-valued equations.

Next, we consider the shear associated with the null congruence along
the outgoing null vector field $l^a$. This quantity is defined as
\begin{equation}
\sigma^{(l)}_{ab} = \left(\gamma_a{}^c\gamma_b{}^d 
 - \frac{1}{2}\, \gamma_{ab}\gamma^{cd} \right) \nabla_c l_d\; ,
\label{Eq:shear}
\end{equation}
where $\gamma_{ab} = g_{ab} + n_a n_b - s_a s_b = 2 m_{(a}
\bar{m}_{b)}$ is the induced metric on $S_t$. Notice that
$\sigma^{(l)}_{ab}$ is normal to $n^a$ and $s^a$ and trace-free; hence
it has two degrees of freedom. Furthermore, it depends on first
derivatives of the metric. We impose the boundary condition
\begin{equation}
m^a m^b \sigma^{(l)}_{ab} \hateq q_2\; ,
\label{Eq:FirstOrderRadiativeBC}
\end{equation}
where $q_2$ is a given complex-valued function on ${\cal T}$. In order
to express this condition in terms of $C^c{}_{ab}$ and background
quantities, we first notice that the one-forms $n_a$ and $s_a$ are
related to their corresponding background quantities $\nz_a$ and
$\sz_a$ by
\begin{displaymath}
n_a = \alpha\, \nz_a\; ,\qquad
s_a = \epsilon\, \sz_a + \delta\, \nz_a\; ,
\end{displaymath}
where $\alpha$, $\delta$ and $\epsilon$ are functions on ${\cal T}$
with $\alpha$ and $\epsilon$ being strictly positive. Next, we compute
\begin{displaymath}
\gamma_a{}^c\gamma_b{}^d \nabla_c l_d 
 = \gamma_a{}^c\gamma_b{}^d \left( \nablaz_c l_d - l_e C^e{}_{cd} \right).
\end{displaymath}
Since $\{ n_a\, , s_a \}$ and $\{ \nz_a\, ,\sz_a \}$ span the same
vector space, $\gamma_a{}^c\gamma_b{}^d \nablaz_c l_d =:
\CKz^{(l)}_{ab}$ coincides with the second fundamental form of the
two-surface $S_t$ as embedded in the {\em background} manifold
$(M,\gz_{ab})$ with respect to the normal vector field $\gz^{ab}
l_b$. Therefore,
\begin{displaymath}
\sigma^{(l)}_{ab} = \left(\gamma_a{}^c\gamma_b{}^d 
 - \frac{1}{2}\, \gamma_{ab}\gamma^{cd} \right)
\left( \CKz^{(l)}_{cd} - l_e C^e{}_{cd} \right).
\end{displaymath}
Since $\sqrt{2}\, l_a = (\alpha + \delta)\nz_a + \epsilon\, \sz_a$,
$\CKz^{(l)}_{ab}$ is explicitly given by
\begin{displaymath}
\CKz^{(l)}_{ab} =  \frac{1}{\sqrt{2}}\gamma_a{}^c\gamma_b{}^d 
   \left[ (\alpha+\delta) \nablaz_c\nz_d + \epsilon\, \nablaz_c\sz_d \right].
\end{displaymath}
For the following, it is important to notice that while
$\CKz^{(l)}_{ab}$ depends on the metric fields $h_{ab}$, it does {\em
not} depend on derivatives of $h_{ab}$. Finally, using
Eq. (\ref{Eq:Christoffel}), the boundary condition
(\ref{Eq:FirstOrderRadiativeBC}) can be expressed as
\begin{equation}
D_{lmm} - 2D_{mlm} \hateq 2\left( q_2 - \CKz^{(l)}_{mm} \right).
\label{Eq:FirstOrderBC4}
\end{equation}

Considering that $\sqrt{2}\, l^a\partial_a = n^a\partial_a +
s^a\partial_a$, we see that the boundary conditions
(\ref{Eq:FirstOrderBC1})-(\ref{Eq:FirstOrderBC3}) and
(\ref{Eq:FirstOrderBC4}) yield generalized Sommerfeld conditions of
the form $l^a\nablaz_a u \hateq q$ for the metric components $u \in \{
h_{m\bar{m}}\, ,h_{kk}\, ,h_{km}\, ,h_{mm} \}$ where $q$ does not
contain any derivatives along $l^a$. For this reason, we specify four
more Sommerfeld-like conditions on the remaining metric components
$h_{ll}$, $h_{lk}$, $h_{lm}$ and obtain the first order boundary
conditions
\begin{eqnarray}
D_{lll} &\hateq & p\;,
\label{Eq:FirstOrderBCList1} \\
D_{llk} &\hateq & \pi\;,
\label{Eq:FirstOrderBCList2} \\
D_{llm} &\hateq & q_1\;,
\label{Eq:FirstOrderBCList3} \\
D_{lmm} &\hateq & 2D_{mlm} + 2\left( q_2 - \CKz^{(l)}_{mm} \right)\,, 
\label{Eq:FirstOrderBCList4} \\
D_{lm\bar{m}} &\hateq &-D_{kll} + D_{ml\bar{m}} + D_{\bar{m}lm} - H_l\;, 
\label{Eq:FirstOrderBCList5} \\
D_{lkm} &\hateq &  -D_{klm} + D_{mlk} + D_{\bar{m}mm} - H_m\;, 
\label{Eq:FirstOrderBCList6} \\
D_{lkk} &\hateq & -D_{km\bar{m}} + D_{m\bar{m}k} + D_{\bar{m}mk} 
- H_k\;,
\label{Eq:FirstOrderBCList7}
\end{eqnarray}
where $p$ and $\pi$ are real-valued given functions on ${\cal T}$ and
$q_1$ and $q_2$ are complex-value given functions on ${\cal T}$, with
$q_2 = m^a m^b\sigma^{(l)}_{ab}$ representing the shear with respect
to the outgoing null vector field $l^a$. With respect to a rotation $m
\mapsto e^{i\varphi} m$ of the complex null vector, we have $q_1
\mapsto e^{i\varphi} q_1$ and $q_2 \mapsto e^{2i\varphi} q_2$; hence
$q_1$ and $q_2$ have spin weights $1$ and $2$, respectively.

%%%%%%%%%%%%%%%%%%%%%%%%%%%%%%%%%%%%%%%%%%%%%%%%%%%%%%%%%%%%%%
\subsection{Second order boundary conditions}
\label{Sect:OuterBC_second}
%%%%%%%%%%%%%%%%%%%%%%%%%%%%%%%%%%%%%%%%%%%%%%%%%%%%%%%%%%%%%%

Next, we generalize the previous boundary conditions to second order
boundary conditions, which depend on second derivatives of
$h_{ab}$. The motivation for this is that such conditions can be used
to reduce the amount of spurious reflections at the boundary
surface. For example, the four boundary conditions
(\ref{Eq:FirstOrderCPBC}) are Dirichlet conditions for the constraint
propagation system, Eq. (\ref{Eq:Bianchi}) with ${\cal E}_{ab} = 0$,
which means that constraint violations are reflected at the boundary
\cite{Rinne:2006vv}. Such reflections can be reduced by replacing the
four Dirichlet conditions with Sommerfeld-like boundary conditions on
the constraint variables ${\cal C}_c$, namely
\begin{equation}
l^a\nabla_a {\cal C}_b \hateq 0\,.
\label{Eq:SecondOrderCPBC}
\end{equation}
These conditions were used in \cite{Lindblom06} and analyzed in
\cite{Rinne:2006vv}. As before, they imply that solutions to the
constraint propagation system with trivial initial data vanish
identically. But in contrast to the condition
(\ref{Eq:FirstOrderCPBC}), the condition (\ref{Eq:SecondOrderCPBC})
allows most constraint violations generated inside the computational
domain by numerical errors to leave the computational
domain.\footnote{One still get reflections for plane waves with
non-normal incidence, for instance. See Sec.~\ref{Sect:Quality} for
more details.} In view of Eq. (\ref{Eq:GradC}) these conditions yield
\begin{eqnarray}
0 \hateq l^a l^b \nabla_a {\cal C}_b &\hateq&
 - E_{llm\bar{m}} - E_{lkll} + E_{lm\bar{m}l} + E_{l\bar{m}ml}
\nonumber\\
& -& 2\,C^{cd}{}_l C_{lcd} - C^c{}_{ll} C_{cef} g^{ef} 
- l^a l^b \nabla_a H_b\; ,
\label{Eq:SecondOrderBC1}\\
0 \hateq l^a k^b \nabla_a {\cal C}_b &\hateq& 
 - E_{llkk} - E_{lkm\bar{m}} + E_{lm\bar{m}k} + E_{l\bar{m}mk} 
\nonumber\\
 &-& 2\,C^{cd}{}_l C_{kcd} - C^c{}_{lk} C_{cef} g^{ef} 
 - l^a k^b \nabla_a H_b\; ,
\label{Eq:SecondOrderBC2}\\
0 \hateq l^a m^b \nabla_a {\cal C}_b &\hateq& 
 - E_{llkm} - E_{lklm} + E_{lmlk} + E_{l\bar{m}mm} 
\nonumber\\
&-& 2\,C^{cd}{}_l C_{mcd} - C^c{}_{lm} C_{cef} g^{ef} 
 - l^a m^b \nabla_a H_b\; ,
\label{Eq:SecondOrderBC3}
\end{eqnarray}
where we have defined the tensor field $E_{cdab} \equiv
\nablaz_c\nablaz_d h_{ab}$.

Next, we specify the complex Weyl scalar $\Psi_0$ at the
boundary. Such boundary conditions have been proposed in the
literature \cite{Friedrich99, Lindblom05, Sarbach:2004rv, Lindblom06,
Rinne:2006vv, Nagy:2006pr, Bardeen:2001xx, Rinne:2007ui}. 
In particular, freezing $\Psi_0$ to its
initial value has been suggested as a good starting point for
absorbing gravitational waves which propagate out of the computational
domain. Recently, an analytic study \cite{Buchman:2006xf} has shown
that this freezing-$\Psi_0$ condition yields spurious reflections
which decay as fast as $(k R)^{-4}$ for large $k R$, for monochromatic
radiation with wavenumber $k$ and for an outer boundary with areal
radius $R$. The Weyl scalar $\Psi_0$ is defined as
\begin{displaymath}
\Psi_0 = R_{abcd}\, l^a\, m^b\, l^c\, m^d.
\end{displaymath}
Using the expression (\ref{Eq:Riemann}) for the Riemann curvature
tensor, we obtain
\begin{displaymath}
2\,\Psi_0 = -E_{llmm} - E_{mmll} + 2E_{(lm)lm}
 + 2\,C^c{}_{lm} C_{clm} 
- 2\,C^c{}_{ll} C_{cmm}
 + l_a\Rz^a{}_{mlm} - m_a\Rz^a{}_{llm}\; .
\end{displaymath}
Therefore, the six boundary conditions specified so far have the form
$l^a l^b\nabla_a\nabla_b u \hateq q$, where $u \in \{
h_{m\bar{m}},h_{kk}, h_{km}, h_{mm} \}$, and $q$ depends on zeroth,
first and second derivatives of $h_{ab}$ but only contains up to
first-order derivatives with respect to $l^a$. We supplement these
conditions with four similar conditions on the missing metric
components $h_{ll}$, $h_{lk}$ and $h_{lm}$. The second order boundary
conditions are then
\begin{eqnarray}
E_{llll} &\hateq& p\; ,
\label{Eq:SecondOrderBCList1} \\
E_{lllk} &\hateq& \pi\; ,
\\
E_{lllm} &\hateq& q_1\; ,
\label{Eq:SecondOrderBCList3}\\
E_{llmm} &\hateq& - E_{mmll} + 2\,E_{(lm)lm}
+ 2\,C^c{}_{lm} C_{clm} 
\nonumber\\
&-& 2C^c{}_{ll} C_{cmm}
+ l_a\Rz^a{}_{mlm} - m_a\Rz^a{}_{llm} 
- 2\psi_0\; , 
\label{Eq:SecondOrderBCList4} \\
E_{llkm} &\hateq& - E_{lklm} + E_{lmlk} + E_{l\bar{m}mm} 
- 2\,C^{cd}{}_l\, C_{mcd} 
\nonumber\\
&-& C^c{}_{lm} C_{cef} g^{ef} 
- l^a m^b \nabla_a H_b\; ,
\label{Eq:SecondOrderBCList5} \\
E_{llm\bar{m}} &\hateq& - E_{lkll} + E_{lm\bar{m}l} + E_{l\bar{m}ml}
 - 2\,C^{cd}{}_l\, C_{lcd} 
\nonumber\\
&-& C^c{}_{ll}\, C_{cef} g^{ef} 
 - l^a l^b \nabla_a H_b\; ,
\label{Eq:SecondOrderBCList6}\\
E_{llkk} &\hateq& - E_{lkm\bar{m}} + E_{lm\bar{m}k} + E_{l\bar{m}mk} 
 - 2\,C^{cd}{}_l\, C_{kcd}
\nonumber\\
& -& C^c{}_{lk} C_{cef} g^{ef} 
 - l^a k^b \nabla_a H_b\; , 
\label{Eq:SecondOrderBCList7}
\end{eqnarray}
where $p$ and $\pi$ are real-valued given functions on ${\cal T}$, and
$q_1$ and $\psi_0$ are complex-valued given functions on ${\cal T}$,
with $\psi_0$ representing the Weyl scalar $\Psi_0$ with respect to
the Newman-Penrose null tetrad constructed in
Eqs. (\ref{Eq:NPTetrad1}--\ref{Eq:NPTetrad2}).

%%%%%%%%%%%%%%%%%%%%%%%%%%%%%%%%%%%%%%%%%%%%%%%%%%%%%%%%%%%%%%
\subsection{Higher order boundary conditions}
\label{Sect:OuterBC_third}
%%%%%%%%%%%%%%%%%%%%%%%%%%%%%%%%%%%%%%%%%%%%%%%%%%%%%%%%%%%%%%

The first and second order boundary conditions constructed so far can
be generalized to arbitrarily high order. Let $L \geq 1$ and consider
the following $(L+1)$ order boundary conditions,
\begin{eqnarray}
&& l^{a_1} l^{a_2} ... l^{a_{L+1}} l^c l^d 
   \nablaz_{a_1}\nablaz_{a_2} ... \nablaz_{a_{L+1}} h_{cd} \hateq p\,,
\label{Eq:HigherOrderGaugeBC1}\\
&& l^{a_1} l^{a_2} ... l^{a_{L+1}} l^c k^d 
   \nablaz_{a_1}\nablaz_{a_2} ... \nablaz_{a_{L+1}} h_{cd} \hateq \pi\,,
\label{Eq:HigherOrderGaugeBC2}\\
&& l^{a_1} l^{a_2} ... l^{a_{L+1}} l^c m^d 
   \nablaz_{a_1}\nablaz_{a_2} ... \nablaz_{a_{L+1}} h_{cd} \hateq q_1\,,
\label{Eq:HigherOrderGaugeBC3}
\end{eqnarray}
together with the four constraint-preserving boundary conditions
\begin{equation}
l^{a_1} l^{a_2} ... l^{a_L} 
 \nabla_{a_1}\nabla_{a_2} ... \nabla_{a_L} {\cal C}_b \hateq 0\,,
\label{Eq:HigherOrderCPBC}
\end{equation}
and the two real-valued boundary conditions
\begin{equation}
l^{a_1} l^{a_2} ... l^{a_{L-1}} l^c m^d l^e m^f 
 \nabla_{a_1}\nabla_{a_2} ... \nabla_{a_{L-1}} C_{cdef} 
 + b(h_{cd}, \nabla_{a} h_{cd}, ... ,\nabla_{a_1}\nabla_{a_2}...\nabla_{a_L}
     h_{cd}; l^a,k^a,m^a)
\hateq 0\,,
\label{Eq:HigherOrderPsi0}
\end{equation}
where the function $b$ depends smoothly on $h_{cd}$, its derivatives
of order smaller than or equal to $L$, and the tetrad vectors $l^a$,
$k^a$ and $m^a$.

Boundary conditions of the form (\ref{Eq:HigherOrderPsi0}) have
recently been constructed in \cite{Buchman:2006xf,Buchman:2007pj}. In
\cite{Buchman:2006xf}, a hierarchy of boundary conditions ${\cal B}_L$
of this form was introduced that lead to perfect absorption for weak
gravitational waves with angular momentum number smaller than or equal
to $L$. This hierarchy was refined in \cite{Buchman:2007pj} where
correction terms from the curvature of the spacetime were taken into
account. Furthermore, as analyzed in Sec.~\ref{Sect:Quality}, the
conditions (\ref{Eq:HigherOrderCPBC}) should yield fewer and fewer
spurious reflections of constraint violations as $L$ is increased. The
geometric meaning of the boundary conditions
(\ref{Eq:HigherOrderGaugeBC1},\ref{Eq:HigherOrderGaugeBC2},\ref{Eq:HigherOrderGaugeBC3})
is not clear. However, their importance lies in the fact that together
with the conditions (\ref{Eq:HigherOrderCPBC}) and
(\ref{Eq:HigherOrderPsi0}) they yield a well posed IBVP in the limit
of frozen coefficients as we shall show in the next section.

We end this section by analyzing how the first order condition
(\ref{Eq:FirstOrderRadiativeBC}) on the shear fits into the hierarchy
(\ref{Eq:HigherOrderPsi0}). For this, consider the Newman-Penrose
field equation
\begin{equation}
  l^a \nabla_a \sigma - m^a \nabla_a \kappa = -\Psi_0 + 
  (\textrm{terms quadratic in the first derivatives of } h_{ab})\,.
\label{Eq:NPEq}
\end{equation}
Here $\sigma = m^a m^b \nabla_b l_a = m^a m^b\sigma_{ab}^{(l)}$ is the
shear, and the spin coefficient $\kappa = m^a l^b \nabla_b l_a$ can be
written as
\begin{displaymath}
  \kappa = -\half D_{mll} + (\textrm{undifferentiated terms in } h_{ab})\,.
\end{displaymath}
If one could impose the boundary condition
that the derivative along $m^a$ of $\kappa$ cancel the quadratic
terms on the right-hand side of Eq. (\ref{Eq:NPEq}), one would obtain
\begin{equation}
  \Psi_0 = -l^a \nabla_a \sigma\,,
\label{Eq:Psi0Limit}
\end{equation}
so that the boundary condition (\ref{Eq:FirstOrderRadiativeBC}) could
be thought of as the ``$L = 0$ member'' of
(\ref{Eq:HigherOrderPsi0}). We have not found a way of achieving this
cancellation for the general case. However, in the high-frequency limit
considered in Sec.~\ref{Sect:ReflCoeffHighFreq} we show that it is
possible to choose coordinates such that the condition $l^a h_{ab} =
0$ is satisfied everywhere and at all times such that $D_{mll} = 0$.
Since in the high-frequency limit the quadratic terms in
Eq. (\ref{Eq:NPEq}) can be neglected, (\ref{Eq:NPEq}) then reduces to
Eq. (\ref{Eq:Psi0Limit}).

%%%%%%%%%%%%%%%%%%%%%%%%%%%%%%%%%%%%%%%%%%%%%%%%%%%%%%%%%%%%%%
\section{Well posedness}
\label{Sect:WellPosedness}
%%%%%%%%%%%%%%%%%%%%%%%%%%%%%%%%%%%%%%%%%%%%%%%%%%%%%%%%%%%%%%

In this section, we analyze the well posedness of the IBVP resulting
from the evolution equations (\ref{Eq:DefE}) with ${\cal E}_{ab} = 0$
with either the first, the second or the higher order boundary
conditions discussed in the previous section. We also consider mixed
first-order second-order boundary conditions very similar to the ones
used in \cite{Lindblom06,Rinne:2006vv}. In order to do so, we use the
frozen coefficient approximation, in which one considers small
amplitude, high frequency perturbations of a given, smooth background
solution \cite{Kreiss89,Gustafsson95}. Intuitively, this is the regime
that is important for the continuous dependence on the data, so it is
expected that if the problem is well posed in the frozen coefficient
approximation, it is also well posed in the full nonlinear case. Using
the theory of pseudo-differential operators and the symmetrizer
construction below to estimate derivatives of arbitrary high order it
should be possible to prove well posedness in the nonlinear case as
well.

For small amplitude, high frequency perturbations of a background
solution (which we take to be $\gz_{ab}$), the evolution equations
(\ref{Eq:DefE}) with ${\cal E}_{ab} = 0$ near a given point $p$ of the
manifold $M$ reduce to
\begin{displaymath}
\gz^{cd}(p) \partial_c\partial_d h_{ab} 
 = 2\partial_{(a} H_{b)} \equiv {\cal F}_{ab}\; ,
\end{displaymath}
where $\gz_{ab}(p)$ is the constant metric tensor obtained from
freezing $\gz$ at the point $p$. Furthermore, the boundary can be
considered to be a plane in our approximation. Therefore, the
nonlinear wave equation is reduced to a linear constant coefficient
problem on the spacetime manifold $\Omega = (0,\infty) \times \Sigma$,
where $\Sigma = \{ (x,y,z)\in \Real^3 : x > 0 \}$ is the half
space. By performing a suitable coordinate transformation which leaves
the foliation $\Sigma_t = \{ t \} \times \Sigma$ invariant, it is
possible to bring the constant metric $\gz_{ab}(p)$ to the simple form
\begin{equation}
\gz(p) = -dt^2 + (dx + \beta\, dt)^2 + dy^2 + dz^2\,,
\label{Eq:FrozenMetric}
\end{equation}
with $\beta$ a constant. In order to see this, assume that $\gz(p)$ is
given by
\begin{displaymath}
\gz(p) = -\alpha^2 dT^2 + h_{ij} (dX^i + \beta^i dT) (dX^j + \beta^j dT)\,,
\end{displaymath}
where $\alpha$ is a positive constant, $\beta^i$ a constant vector,
and $h_{ij} dX^i dX^j = N^2 (dX^1)^2 + H_{AB}(dX^A + b^A dX^1)(dX^B +
b^B dX^1)$ is a constant, positive definite three-metric. Here, $A,B$
are equal to $2$ or $3$, $N$ is a positive constant, $b^A$ a constant
two-vector and $H_{AB}$ a positive definite two-metric. Then, a
suitable change of the coordinates $X^2$ and $X^3$ gives $H_{AB} =
\delta_{AB}$. Next, the transformation $Y^1 = N X^1$, $Y^A = X^A + b^A
X^1$ leaves the domain $\Sigma$ invariant and brings the three metric
into the form $h_{ij} = \delta_{ij}$. Finally, we perform the
transformation $t = \alpha T$, $x = Y^1$, $y = Y^2 + \beta^2 T$, $z =
Y^3 + \beta^3 T$, which leaves the foliation $\Sigma_t$ invariant and
brings the metric into the form (\ref{Eq:FrozenMetric}). With respect
to this metric, the evolution equations reduce to
\begin{equation}
\left[ -\partial_t^2 + 2\beta\partial_t\partial_x + (1-\beta^2)\partial_x^2
 + \partial_y^2 +\partial_z^2 \right] h_{ab} 
 = {\cal F}_{ab}\; .
\label{Eq:FrozenCoeffWaveEq}
\end{equation}
For the following, we assume that the constant $\beta$ is smaller than
one in magnitude. Although this condition might not hold everywhere if
black holes are present, it holds near the boundary since the boundary
surface $\mathcal{T}$ is assumed to be time-like.

%%%%%%%%%%%%%%%%%%%%%%%%%%%%%%%%%%%%%%%%%%%%%%%%%%%%%%%%%%%%%%
\subsection{First order boundary conditions}
\label{Sect:WellPosedness_first}
%%%%%%%%%%%%%%%%%%%%%%%%%%%%%%%%%%%%%%%%%%%%%%%%%%%%%%%%%%%%%%

With respect to the metric (\ref{Eq:FrozenMetric}), we have
\begin{displaymath}
\nz^a\partial_a = \partial_t - \beta\partial_x\; ,\qquad
\sz^a\partial_a = -\partial_x\,,
\end{displaymath}
and therefore, an adapted {\em background} null tetrad to $S_t$ is
\begin{eqnarray}
l^a\partial_a &=& \frac{1}{\sqrt{2}}\left[ 
 \partial_t - (1 + \beta)\partial_x \right], \qquad
\nonumber\\
k^a\partial_a &=& \frac{1}{\sqrt{2}}\left[ 
 \partial_t + (1 - \beta)\partial_x \right], \qquad
\label{eq:tetrad}\\
m^a\partial_a &=& \frac{1}{\sqrt{2}}\left[ \partial_y + i\partial_z \right].
\nonumber
\end{eqnarray}
In the frozen coefficient approximation, the first order boundary
conditions (\ref{Eq:FirstOrderBCList1}--\ref{Eq:FirstOrderBCList7})
reduce to
\begin{eqnarray}
l^a\partial_a h_{ll} &\hateq& p\; ,
\label{Eq:FrozenFOBC1}\\
l^a\partial_a h_{lk} &\hateq& \pi\; ,
\label{Eq:FrozenFOBC2}\\
l^a\partial_a h_{lm} &\hateq& q_1\; ,
\label{Eq:FrozenFOBC3}\\
%%%%%%%%%%%%%%%%%%%%%%%%%%%%%%%%%%%%%%%%%%%%%%%%%%%%%%%%%%%%%%%%%%%%
l^a\partial_a h_{mm} &\hateq& 2\, m^a\partial_a h_{lm} 
                            + 2\,\left( q_2 - \CKz^{(l)}_{mm} \right),
\label{Eq:FrozenFOBC4}\\
%%%%%%%%%%%%%%%%%%%%%%%%%%%%%%%%%%%%%%%%%%%%%%%%%%%%%%%%%%%%%%%%%%%%
l^a\partial_a h_{m\bar{m}} &\hateq&  
 -k^a\partial_a h_{ll} + m^a\partial_a h_{l\bar{m}} 
 + \bar{m}^a\partial_a h_{lm} 
- H_l\; ,
\label{Eq:FrozenFOBC5}\\
l^a\partial_a h_{km} &\hateq& 
 -k^a\partial_a h_{lm} + m^a\partial_a h_{lk} 
 + \bar{m}^a\partial_a h_{mm} 
- H_m\; ,
\label{Eq:FrozenFOBC6}\\
l^a\partial_a h_{kk} &\hateq& 
 -k^a\partial_a h_{m\bar{m}} + m^a\partial_a h_{\bar{m}k} 
 + \bar{m}^a\partial_a h_{mk} 
- H_k\; .
\label{Eq:FrozenFOBC7}
\end{eqnarray}
Notice that the derivatives along $k^a$ in
Eqs. (\ref{Eq:FrozenFOBC5}) and (\ref{Eq:FrozenFOBC6}) could be replaced by
derivatives along the time-evolution vector field $\partial_t$ by
using $k^a\partial_a = \left[ \sqrt{2}\partial_t - (1-\beta)
l^a\partial_a \right]/(1+\beta)$ and
Eqs. (\ref{Eq:FrozenFOBC1}) and (\ref{Eq:FrozenFOBC3}). Similarly, the term
$k^a\partial_a h_{m\bar{m}}$ in Eq. (\ref{Eq:FrozenFOBC7}) can be
replaced by tangential derivatives by using $k^a\partial_a = \left[
\sqrt{2}\partial_t - (1-\beta) l^a\partial_a \right]/(1+\beta)$ and
the new version of Eq. (\ref{Eq:FrozenFOBC5}). In this way, only
derivatives tangential to the boundary appear on the right-hand sides
of Eqs. (\ref{Eq:FrozenFOBC1}--\ref{Eq:FrozenFOBC7}). While this
observation might be useful for numerical work it is not important for
what follows. The evolution system
(\ref{Eq:FrozenCoeffWaveEq},\ref{Eq:FrozenFOBC1}--\ref{Eq:FrozenFOBC7})
has the form of a cascade of wave problems of the form
\begin{eqnarray}
&&\left[ -\partial_t^2 + 2\beta\partial_t\partial_x + (1-\beta^2)\partial_x^2
 + \partial_y^2 + \partial_z^2 \right] u^{(i)} 
= {\cal F}^{(i)}\,,
\hspace{1cm} \hbox{on $\Omega$}\,,
\label{Eq:WaveProblemEq}\\
&&\left[ \partial_t - (1 + \beta)\partial_x \right] u^{(i)} 
\hateq q^{(i)}\,,\hspace{.5cm}\hbox{on ${\cal T}$}\,,
\label{Eq:WaveProblemBC}
\end{eqnarray}
where $i=1,...,10$ and where the boundary data $q_i$ depends on {\em
first order derivatives of the fields $u^{(j)}$, for $j=1,...,i-1$
only, at the boundary surface ${\cal T}$}.

In the following, we obtain a priori estimates for each wave problem, Eqs.
(\ref{Eq:WaveProblemEq}) and (\ref{Eq:WaveProblemBC}), using the method in
\cite{Kreiss:2006mi}. For this, we first remark that it is sufficient
to consider the case of trivial initial data. Indeed, let $u =
u^{(i)}$ be any smooth solution to
(\ref{Eq:WaveProblemEq}--\ref{Eq:WaveProblemBC}). Then,
\begin{equation}
\bar{u}(t,x,y,z) = u(t,x,y,z) - u(0,x,y,z) - t\,\partial_t u(0,x,y,z)\,,
\qquad t \geq 0\,, \qquad (x,y,z)\in\Sigma\,,
\label{Eq:TransToTrivialID}
\end{equation}
satisfies the wave problem
(\ref{Eq:WaveProblemEq}--\ref{Eq:WaveProblemBC}) with modified source
functions ${\cal F}^{(i)}$ and $q^{(i)}$ and trivial initial data
$\bar{u}(0,x,y,z) = 0$, $\partial_t\bar{u}(0,x,y,z) = 0$ for all
$(x,y,z)\in \Sigma$. Next, we show there exists a constant $C_i > 0$
such that for all $\eta > 0$ and all smooth enough solutions $u^{(i)}$
of (\ref{Eq:WaveProblemEq}--\ref{Eq:WaveProblemBC}) with trivial
initial data,
\begin{equation}
\eta\, \| u^{(i)} \|^2_{\eta,1,\Omega} 
   + \| u^{(i)} \|^2_{\eta,1,{\cal T}}
\leq C_i\,\left( \eta^{-1} \| {\cal F}^{(i)} \|^2_{\eta,0,\Omega} 
  + \| q^{(i)} \|^2_{\eta,0,{\cal T}} \right)\,,
\label{Eq:WaveEstimate}
\end{equation}
where the norms above are defined as
\begin{eqnarray}
\| u \|^2_{\eta,m,\Omega} &=&
\int_{\Omega} e^{-2\eta t}
\sum\limits_{|\alpha|\leq m}
| \partial_t^{\alpha_t}\,\partial_x^{\alpha_x}
\partial_y^{\alpha_y}\,
\partial_z^{\alpha_z} u(t,x,y,z) |^2\,dt\, dx\, dy\, dz\,,
\nonumber\\
\| u \|^2_{\eta,m,{\cal T}} &=& \int_{\cal T} e^{-2\eta t}
\sum\limits_{|\alpha|\leq m} |\partial_t^{\alpha_t}\,\partial_x^{\alpha_x}
\partial_y^{\alpha_y}\,\partial_z^{\alpha_z} u(t,0,y,z) |^2 dt\, dy\, dz\,,
\nonumber
\end{eqnarray}
where $\alpha = (\alpha_t,\alpha_x,\alpha_y,\alpha_z)\in \Natural_0^4$
is a multi-index and $|\alpha| = \alpha_t + \alpha_x + \alpha_y +
\alpha_z$. The important point to notice here is that one obtains an
estimate for the $L^2$ norm of the {\em first order derivatives of the
solution with respect to the boundary surface ${\cal T}$}. Therefore,
in the estimate of the $i$'th wave problem, the norms of the first
derivatives of the fields $u^{(j)}$, $j=1,...,i-1$, at the boundary
which appear in the norm of $q^{(i)}$ on the right-hand side of
(\ref{Eq:WaveEstimate}) can be estimated and one obtains the following
global estimate.\footnote{Problems which satisfy this kind of estimate
together with existence of solutions are called {\em strongly well
posed in the generalized sense} in the literature
\cite{Kreiss89,Kreiss:2006mi}. Here, ``generalized sense'' refers to
the fact that trivial initial data is assumed and that the norms
involve a time integration. As illustrated above, the assumption of
trivial initial data does not restrict the solution space since one
can always satisfy it by means of a transformation of the type
(\ref{Eq:TransToTrivialID}), provided the data is sufficiently
smooth. However, since this transformation introduces third
derivatives of the initial data into the source terms ${\cal F}_{ab}$,
it is not clear if our results can be strengthened to obtain {\em
strong well posedness} \cite{Kreiss89}, which does not assume trivial
initial data and where the norms do not contain a time integral.}
There is a constant $C > 0$ such that for all $\eta > 0$ and smooth
enough solutions $h_{ab}$ of the initial-boundary value problem
(\ref{Eq:FrozenCoeffWaveEq},\ref{Eq:FrozenFOBC1}--\ref{Eq:FrozenFOBC7})
with trivial initial data,
\begin{eqnarray}
\sum\limits_{a,b=0}^3 \left( 
 \eta \| h_{ab} \|^2_{\eta,1\Omega} + \| h_{ab} \|^2_{\eta,1,{\cal T}} \right)
&\leq& C
\left( \eta^{-1} \sum\limits_{a,b=0}^3 
       \| {\cal F}_{ab} \|^2_{\eta,0,\Omega} 
     + \| p \|^2_{\eta,0,{\cal T}}  + \| \pi \|^2_{\eta,0,{\cal T}}\right.
\nonumber\\
&&\left.    +\; \| q_1 \|^2_{\eta,0,{\cal T}}  + \| q_2 \|^2_{\eta,0,{\cal T}} 
     + \sum\limits_{a=0}^3 \| H_a \|^2_{\eta,0,{\cal T}} \right)\,.
\nonumber
\end{eqnarray}

In order to prove the estimates (\ref{Eq:WaveEstimate}), let $u =
u^{(i)}$ be a solution of one of the wave problems
(\ref{Eq:WaveProblemEq}--\ref{Eq:WaveProblemBC}) with trivial initial
data, i.e. $u(0,x,y,z) = 0$, $\partial_t u(0,x,y,z) = 0$ for
$(x,y,z)\in \Sigma$. Next, fix $\eta > 0$ and define
\begin{eqnarray}
u_\eta(t,x,y,z) =
\left\{ \begin{array}{ll} 
 e^{-\eta t}\,u(x,y,z)  & \hbox{for $t > 0$, $(x,y,z)\in\Sigma$\,,} \\ 
 0                      & \hbox{for $t \leq 0$, $(x,y,z) \in\Sigma$\,.}
\end{array}\, \right.
\label{Eq:ueta}
\end{eqnarray}
Let $\tilde{u}_\eta(\xi,x,\omega_y,\omega_z)$ denote the Fourier
transformation of $u_\eta(t,x,y,z)$ with respect to the directions
$t$, $y$ and $z$ tangential to the boundary, and let
$\tilde{u}(s,x,\omega_y,\omega_z) =
\tilde{u}_\eta(\xi,x,\omega_y,\omega_z)$, $s = \eta + i\xi$, denote
the Fourier-Laplace transformation of $u$. Then, $\tilde{u}$
satisfies the ordinary differential system
\begin{eqnarray}
&&\left[ -s^2 + 2\beta\,s\,\partial_x 
  + (1-\beta^2)\,\partial_x^2 - \omega^2 \right]
 \tilde{u} = 
\tilde{F}\,,
\hspace{0.5cm}
\hbox{on $x \in (0,\infty)$},\\
&&\left[ s - (1 + \beta)\,\partial_x \right]\tilde{u} \hateq \tilde{q}
\hspace{0.5cm}
\hbox{at $x=0$}\,,
\end{eqnarray}
where $\omega = \sqrt{\omega_y^2 + \omega_z^2}$ and $\tilde{F}$ and
$\tilde{q}$ denote the Fourier-Laplace transformations of ${\cal
F}^{(i)}$ and $q^{(i)}$, respectively. We rewrite this as a first
order system by introducing the variable
\begin{equation}
\tilde{v} = \frac{1}{k} 
\left( \partial_x + \gamma^2\beta\, s \right)\tilde{u}\,,
\end{equation}
where $k = \sqrt{ |s|^2 + \omega^2 }$ and $\gamma = 1/\sqrt{1 -
\beta^2}$.  With respect to this, the system can be rewritten in the
form
\begin{eqnarray}
\partial_x\tilde{w} = M(s,\omega) \tilde{w} + \tilde{f}\,,
&& \textrm{on } x \in (0,\infty),
\label{Eq:GeneralFormEq}\\
L(s,\omega)\tilde{w} \hateq \tilde{g}\; ,
&& \textrm{at } x = 0\, ,
\label{Eq:GeneralFormBC}
\end{eqnarray}
where
\begin{displaymath}
\tilde{w} = \left( \begin{array}{c} \tilde{u} \\ \tilde{v} \end{array} \right)\,,
\qquad
\tilde{f} = \frac{\gamma^2}{k} 
 \left( \begin{array}{c} 0 \\ \tilde{F} \end{array} \right)\,,
\qquad
\tilde{g} = \frac{1-\beta}{k}\, \tilde{q}\; ,
\end{displaymath}
and
\begin{displaymath}
M(s,\omega) = k\left( \begin{array}{cc}
 -\gamma^2\beta\, s'  & 1 \\
 \gamma^4(s'^2 + \gamma^{-2}\omega'^2) & -\gamma^2\beta\, s'
\end{array} \right), \qquad
L(s,\omega) = \left( s', -\gamma^{-2} \right),
\end{displaymath}
with $s' = s/k$ and $\omega' = \omega/k$. Notice that $|s'|^2 +
|\omega'|^2 = 1$. The eigenvalues and corresponding eigenvectors of
$M$ are given by
\begin{displaymath}
\mu_\pm = \gamma^2 k
\left( -\beta\, s' \pm \sqrt{s'^2 + \gamma^{-2}\omega'^2} \right)\,,
\qquad
e_{\pm} 
 = \left( \begin{array}{c} 1 \\ \pm \gamma^2\sqrt{s'^2 + \gamma^{-2}\omega'^2}
\end{array} \right),
\end{displaymath}
where the square root is defined to have positive real part for
$\re(s') > 0$. One can show \footnote{See Lemma 2 of
Ref. \cite{Kreiss:2006mi} or use the following argument: Let
$\eta',\xi',a,b$ be real numbers such that $s' = \eta' + i\xi'$ and
$\sqrt{s'^2 + \gamma^{-2}\omega'^2} = a + ib$. Taking the square of
the last equation yields $\eta'\xi' = a b$ and $a^4 + (\xi'^2 -
\eta'^2 - \gamma^{-2}\omega'^2)a^2 - \eta'^2\xi'^2 = 0$, from which
one concludes that $a^2 \geq \eta'^2$.} that in this case
$\re(\sqrt{s'^2 + \gamma^{-2}\omega'^2}) \geq \re(s')$ which implies
that $\re(\mu_-) < 0 < \re(\mu_+)$. Therefore, the solution of
(\ref{Eq:GeneralFormEq}--\ref{Eq:GeneralFormBC}) belonging to a
trivial source term, $\tilde{f} = 0$, which decays as $x \to \infty$
is given by
\begin{equation}
\tilde{w}(s,x,\omega) = \sigma\, e^{\mu_- x} e_-\; ,
\end{equation}
where the constant $\sigma$ satisfies $L(s,\omega) e_- \sigma =
\tilde{g}$, i.e.
\begin{equation}
\left[\, s' + \sqrt{s'^2 + \gamma^{-2}\omega'^2}\, \right]\sigma = \tilde{g}\,.
\end{equation}
It can be shown that there is a strictly positive constant $\delta_2 >
0$ such that $|s' + \sqrt{s'^2 + \gamma^{-2}\omega'^2}| \geq \delta_2$
for all $\re(s') > 0$ and all $\omega'\in \Real$ with $|s'|^2 +
|\omega'|^2 = 1$.\footnote{See the proof of Lemma 3 of
Ref. \cite{Kreiss:2006mi} or use the fact that this condition is
equivalent to $\frac{|\zeta + \sqrt{\zeta^2 +
\gamma^{-2}}|}{\sqrt{|\zeta|^2 + 1}} \geq \delta_2$ for all
$\zeta\in\Complex$ with $\re(\zeta) > 0$. If $|\zeta|\to\infty$ the
left-hand side converges to $2$. For finite $\zeta$ this inequality
follows from Lemma 3.1 of Ref. \cite{Reula:2004nr}.}  Therefore, there
is a constant $C_1 > 0$ such that
\begin{equation}
|\tilde{w}(s,0,\omega)| \leq C_1 |\tilde{g}(s,\omega)|\,,
\end{equation}
for all $\re(s) > 0$ and $\omega\in\Real$. According to the
terminology in \cite{Kreiss:2006mi} this means that the system is {\em
boundary stable}. The key result in \cite{Kreiss70,Kreiss:2006mi} is that
this implies the existence of a symmetrizer $H = H(s',\omega')$, where
$H$ is a complex, two by two Hermitian matrix such that
\begin{enumerate}
\item[(i)] $H(s',\omega')$ depends smoothly on $(s',\omega')$.
\item[(ii)] There exists a constant $\varepsilon_1 > 0$ such that
\begin{displaymath}
H M + M^* H \geq \varepsilon_1 \re(s) I_2\; ,
\end{displaymath}
for all $\re(s) > 0$ and all $\omega\in\Real$, where $I_2$ denotes the
two by two identity matrix.
\item[(iii)]
There are constants $\varepsilon_2 > 0$ and $C_2 > 0$ such that
\begin{displaymath}
< \tilde{w}, H\tilde{w} > \; \geq \varepsilon_2 |\tilde{w}|^2 
 - C_2 |\tilde{g}|^2\,,
\end{displaymath}
for all $\tilde{w}$ satisfying the boundary condition
$L(s,\omega)\tilde{w} = \tilde{g}$, where $< . ,. >$ denotes the
standard scalar product on $\Complex^2$ and $| . |$ the corresponding
norm.
\end{enumerate}
Using this symmetrizer, the estimate (\ref{Eq:WaveEstimate}) can be
obtained as follows. First, using Eq. (\ref{Eq:GeneralFormEq}) and
(ii) we have
\begin{eqnarray}
\partial_x < \tilde{w}, H\tilde{w} > 
 &=& 2< \tilde{w}, H\partial_x\tilde{w} >
\nonumber\\
 &=& < \tilde{w}, (H M + M^* H)\tilde{w} > + 2<\tilde{w}, H\tilde{f} >
\nonumber\\
 &\geq& \varepsilon_1 \re(s) |\tilde{w}|^2 
 - K |\tilde{w}|^2 - \frac{1}{K} | H\tilde{f} |^2,
\nonumber
\end{eqnarray}
where $K > 0$. Integrating both sides from $x=0$ to $\infty$ and
choosing $K = \varepsilon_1\re(s)/2$, we obtain, using (iii),
\begin{eqnarray}
&\re(s)&\int\limits_0^\infty |\tilde{w}|^2 dx 
\leq\frac{2}{\varepsilon_1}\left[ 
   -\left. < \tilde{w}, H\tilde{w} > \right|_{x=0} 
   + \frac{2}{\varepsilon_1\re(s)}\int\limits_0^\infty | H\tilde{f} |^2 dx
\right]
\nonumber\\
 &\leq& \frac{2}{\varepsilon_1}\left( 
 -\varepsilon_2 \left. |\tilde{w} |^2 \right|_{x=0} + C_2 |\tilde{g}|^2 \right)
 + \frac{4}{\varepsilon_1^2\re(s)}\int\limits_0^\infty |H\tilde{f}|^2 dx\,.
\nonumber
\end{eqnarray}
Since $H = H(s',\omega')$ depends smoothly on $(s',\omega')$ and
$|s'|^2 + |\omega'|^2 = 1$, there is a constant $C_3 > 0$ such that
$|H\tilde{f}| \leq C_3 |\tilde{f}|$ for all $(s',\omega')$ satisfying
$\re(s') > 0$ and $|s'|^2 + |\omega'|^2 = 1$. Using this and
multiplying the above inequality by $k^2$ on both sides, we obtain
\begin{equation}
\eta \int\limits_0^\infty 
 \left( | k\tilde{u}|^2 + |\partial_x\tilde{u}|^2 \right) dx
+ \left. \left( | k\tilde{u}|^2 + |\partial_x\tilde{u}|^2 \right) \right|_{x=0}
\leq C\left[ 
 \eta^{-1} \int\limits_0^\infty |\tilde{F} |^2 dx + |\tilde{q}|^2 \right]\,,
\label{Eq:FirstWaveEstimate}
\end{equation}
for some constant $C > 0$. The estimate (\ref{Eq:WaveEstimate})
follows from this after integrating over $\xi = \im(s)$, $\omega_y$
and $\omega_z$ and using Parseval's identity. Existence of solutions
follows from Eqs.~(\ref{Eq:GeneralFormEq}--\ref{Eq:GeneralFormBC}) and
standard results on ordinary differential equations.

Before we proceed to the higher order boundary conditions, we remark
that the estimate (\ref{Eq:WaveEstimate}) can be generalized to the
following statement. For each $m=2,3,4,...$ there exists a constant
$C_{i,m}$ such that
\begin{equation}
\eta \| u^{(i)} \|^2_{\eta,m,\Omega} 
   + \| u^{(i)} \|^2_{\eta,m,{\cal T}}
\leq C_{i,m}\,\left( \eta^{-1} \| {\cal F}^{(i)} \|^2_{\eta,m-1,\Omega} 
  + \| {\cal F}^{(i)} \|^2_{\eta,m-2,{\cal T}} 
  + \| q^{(i)} \|^2_{\eta,m-1,{\cal T}} \right),
\label{Eq:WaveEstimateHigher}
\end{equation}
for all $\eta > 0$ and all smooth enough solutions $u^{(i)}$ with the
property that their first $m$ time derivatives vanish identically at
$t=0$. The latter can always be achieved by means of the
transformation
\begin{displaymath}
\bar{u}^{(i)}(t,x,y,z) = u^{(i)}(t,x,y,z)
 - \sum\limits_{k=0}^m \frac{t^k}{k!}\,(\partial_t)^k u^{(i)}(0,x,y,z)\,,
\qquad t \geq 0\,, \qquad (x,y,z)\in \Sigma\,.
\end{displaymath}
Notice that the evolution equations then imply that the first $(m-2)$
time derivatives of $F$ vanish identically at $t=0$. In order to prove
the estimate (\ref{Eq:WaveEstimateHigher}), we first multiply both
sides of (\ref{Eq:FirstWaveEstimate}) by $k^{2j}$,
$j=1,2,...m-1$. This yields the desired estimates for the tangential
derivatives. In order to estimate the normal derivatives, we use the
evolution equation $\partial_x^2\tilde{u} = \gamma^2\left[ (s^2 +
\omega^2)\tilde{u} - 2\beta s\partial_x\tilde{u} + \tilde{F} \right]$
and the fact that $\eta = \re(s) \leq k$ and obtain
\begin{eqnarray}
\int\limits_0^\infty 
\eta \sum\limits_{j=0}^m | k^{m-j}\partial_x^j\tilde{u}|^2 dx
+ \left. 
\sum\limits_{j=0}^m | k^{m-j}\partial_x^j\tilde{u}|^2 \right|_{x=0}
 &\leq& \tilde{C}_m\left[ \int\limits_0^\infty 
  \eta^{-1} \sum\limits_{j=0}^{m-2} | k^{m-1-j}\partial_x^j\tilde{F}|^2 dx
 \right. \nonumber\\
 &+& \left. \left.\sum\limits_{j=0}^{m-2} | k^{m-2-j}\partial_x^j\tilde{F}|^2 
   \right|_{x=0} 
 + |k^{m-1} \tilde{q}|^2 \right]\,,
\nonumber
\end{eqnarray}
for some constant $\tilde{C}_m$. The estimate
(\ref{Eq:WaveEstimateHigher}) then follows after integrating over $\xi
= \im(s)$, $\omega_y$ and $\omega_z$ and using Parseval's identity.

%%%%%%%%%%%%%%%%%%%%%%%%%%%%%%%%%%%%%%%%%%%%%%%%%%%%%%%%%%%%%%
\subsection{Second and higher order boundary conditions}
\label{Sect:WellPosedness_second}
%%%%%%%%%%%%%%%%%%%%%%%%%%%%%%%%%%%%%%%%%%%%%%%%%%%%%%%%%%%%%%

Next, we generalize the previous estimate to boundary conditions of
arbitrary order $m \geq 2$. In the frozen coefficient limit, the
evolution system with the second or higher order boundary conditions
discussed in the previous section has the form
\begin{eqnarray}
&&\left[ -\partial_t^2 + 2\beta\partial_t\partial_x + (1-\beta^2)\partial_x^2
 + \partial_y^2 + \partial_z^2 \right] u^{(i)} 
={\cal F}^{(i)}\,,
\hspace{0.5cm}\hbox{on $\Omega$}\,,
\label{Eq:WaveProblemEqHigher}\\
&&\left[ \partial_t - (1 + \beta)\partial_x \right]^m u^{(i)} \hateq q^{(i)}\,,
\hspace{0.5cm} \hbox{on ${\cal T}$}\,,
\label{Eq:WaveProblemBCHigher}
\end{eqnarray}
where $i=1,2,...10$ and the boundary data $q^{(i)}$ depends on the
$m$'th derivatives of the fields $u^{(j)}$ for $j=1,...,i-1$
only. Assuming trivial initial data, defining $u_\eta$ as in
Eq. (\ref{Eq:ueta}) and taking the Fourier transformation with respect
to the tangential directions $(t,y,z)$, one obtains the same first
order system (\ref{Eq:GeneralFormEq}) as before, but where the
boundary condition (\ref{Eq:GeneralFormBC}) is replaced by
\begin{displaymath}
{\cal L}^m \tilde{u} \hateq 
\left( \frac{1-\beta}{k} \right)^m \tilde{q}\,,
\end{displaymath}
with the linear operator ${\cal L} \equiv (1-\beta)s' - \gamma^{-2} k^{-1}
\partial_x$. In order to rewrite this in algebraic form, we notice
that by virtue of Eq. (\ref{Eq:GeneralFormEq})
\begin{displaymath}
\left( \begin{array}{c} 
 {\cal L} \tilde{u} \\ {\cal L} \tilde{v} 
\end{array} \right) 
 = B\left( \begin{array}{c} \tilde{u} \\ \tilde{v} \end{array} \right) 
 - \frac{1}{k^2} 
   \left( \begin{array}{c} 0 \\ \tilde{F} \end{array} \right),
\end{displaymath}
where the matrix $B$ is given by
\begin{displaymath}
B = \left( \begin{array}{cc} s' & -\gamma^{-2} \\ 
  -\gamma^2\lambda'^2 & s' \end{array} \right),
\end{displaymath}
where $\lambda' = \sqrt{s'^2 + \gamma^{-2}\omega'^2}$ has positive
real part. Iterating, we obtain
\begin{displaymath}
\left( \begin{array}{c} 
 {\cal L}^m \tilde{u} \\ {\cal L}^m \tilde{v} 
\end{array} \right) 
 = B^m
  \left( \begin{array}{c} \tilde{u} \\ \tilde{v} \end{array} \right) 
 - \frac{1}{k^2}\sum\limits_{j=0}^{m-1} B^j 
   \left( \begin{array}{c} 0 \\ {\cal L}^{m-1-j}\tilde{F} 
    \end{array} \right).
\end{displaymath}
Explicitly, one finds
\begin{displaymath}
B^j = \frac{1}{2}\left( \begin{array}{cc} 
 b_+^j + b_-^j & -\gamma^{-2}\lambda'^{-1}(b_+^j - b_-^j) \\ 
  -\gamma^2\lambda'(b_+^j - b_-^j) & b_+^j + b_-^j \end{array} \right),
\end{displaymath}
where $b_{\pm} = s' \pm \lambda'$ are the eigenvalues of the matrix
$B$. Therefore, the boundary conditions can be brought into the form
(\ref{Eq:GeneralFormBC}) with
\begin{displaymath}
L(s,\omega) = \frac{1}{2}\left( b_+^m + b_-^m, 
   -\gamma^{-2}\lambda'^{-1}(b_+^m - b_-^m) \right)\,,
\end{displaymath}
and
\begin{displaymath}
\tilde{g} = \left( \frac{1-\beta}{k} \right)^m \tilde{q}
 - \frac{1}{2\,\gamma^2\, k\,\lambda'} \sum\limits_{j=1}^{m-1}
   (b_+^j - b_-^j){\cal L}^{m-1-j}\left. \tilde{F} \right|_{x=0}\; .
\end{displaymath}
The solution belonging to a trivial source term, $\tilde{f} = 0$,
which decays as $x \to \infty$ is given by
\begin{equation}
\tilde{w}(s,x,\omega) = \sigma e^{\mu_- x} e_-\; ,
\end{equation}
where the constant $\sigma$ satisfies $L(s,\omega) e_- \sigma =
\tilde{g}$. Since $e_- = (1, -\gamma^2\lambda')^T$, this condition
reduces to
\begin{displaymath}
b_+^m\sigma = \tilde{g}\,.
\end{displaymath}
However, as was shown in the last section, there is a constant
$\delta_2 > 0$ such that $|b_+| \geq \delta_2$ for all $\re(s') > 0$
and all $\omega'\in\Real$ with $|s'|^2 + |\omega'|^2 = 1$. Therefore,
there is a constant $C_2 > 0$ such that
\begin{equation}
|\tilde{w}(s,0,\omega)| \leq C_2 |\tilde{g}(s,\omega)|\,,
\end{equation}
for all $\re(s) > 0$ and $\omega\in\Real$ and the system is boundary
stable. Therefore, the exists a smooth symmetrizer satisfying the
conditions (i)--(iii) above and we obtain the estimate
\begin{displaymath}
\eta \int\limits_0^\infty |\tilde{w}|^2 dx 
 + \left. |\tilde{w}|^2\right|_{x=0}
\leq C\left[ \eta^{-1} \int\limits_0^\infty |\tilde{f} |^2 dx 
 + |\tilde{g}|^2 \right]\,,
\end{displaymath}
for some constant $C > 0$. Multiplying both sides by $k^{2m}$, using
the evolution equation $\partial_x^2\tilde{u} = \gamma^2\left[ (s^2 +
\omega^2)\tilde{u} - 2\beta s\partial_x\tilde{u} + \tilde{F} \right]$
and $\eta = \re(s) \leq k$, we obtain the estimate
\begin{eqnarray}
\eta
\int\limits_0^\infty 
\sum\limits_{j=0}^m | k^{m-j}\partial_x^j\tilde{u}|^2 dx
+ \left. 
\sum\limits_{j=0}^m | k^{m-j}\partial_x^j\tilde{u}|^2 \right|_{x=0}
&\leq& \tilde{C}\left[ 
\eta^{-1} \int\limits_0^\infty
  \sum\limits_{j=0}^{m-2} | k^{m-1-j}\partial_x^j\tilde{F}|^2 dx\right.
\nonumber\\
&  +& \left.\left. \sum\limits_{j=0}^{m-2} | k^{m-2-j}\partial_x^j\tilde{F}|^2 
    \right|_{x=0} 
  + |\tilde{q}|^2 \right]\,,
\end{eqnarray}
for some new constant $\tilde{C} > 0$. Using Parseval's relations and
assuming that $\partial_t^j u(0,x,y,z) = 0$ for all $j=0,1,...m$ we
have
\begin{eqnarray}
&\eta& \| u \|_{\eta,m,\Omega}^2 + \| u \|_{\eta,m,{\cal T}}^2 
 \leq \hat{C}\left[ \eta^{-1} \| F \|_{\eta,m-1,\Omega}^2 
+ \| F \|_{\eta,m-2,{\cal T}}^2
+\| q \|_{\eta,0,{\cal T}}^2 \right]\,.
\label{Eq:HigherOrderEstimate}
\end{eqnarray}
Therefore, we obtain an a priori estimate as before.

%%%%%%%%%%%%%%%%%%%%%%%%%%%%%%%%%%%%%%%%%%%%%%%%%%%%%%%%%%%%%%
\subsection{Mixed first and second order boundary conditions}
\label{Sect:WellPosedness_mixed}
%%%%%%%%%%%%%%%%%%%%%%%%%%%%%%%%%%%%%%%%%%%%%%%%%%%%%%%%%%%%%%

In some cases, similar estimates can be proved for combinations of
first order and second order boundary conditions. Here, we consider
the boundary conditions that are obtained by combining the first order
gauge boundary conditions
(\ref{Eq:FirstOrderBCList1}--\ref{Eq:FirstOrderBCList3}) with the
second order constraint-preserving boundary conditions
(\ref{Eq:SecondOrderBCList4}--\ref{Eq:SecondOrderBCList7}) which
specify $\Psi_0$. This set of boundary conditions was used in
\cite{Lindblom06,Rinne:2006vv,Rinne:2007ui}, and we shall also use
them in one of our numerical tests in Sec.~\ref{Sect:NumTests}. As
before, we work in the frozen coefficient approximation.

Consider first the first order gauge boundary conditions
(\ref{Eq:FrozenFOBC1}--\ref{Eq:FrozenFOBC3}). Using the estimate
(\ref{Eq:WaveEstimateHigher}) with $m=2$, we have
\begin{equation}
\sum\limits_{a=0}^3 \left( \eta \| h_{la} \|^2_{\eta,2,\Omega}
 + \| h_{la} \|^2_{\eta,2,{\cal T}} \right)
\leq C_1
\left[ \sum\limits_{a=0}^3 \left( \eta^{-1} 
       \| {\cal F}_{la} \|^2_{\eta,1,\Omega} 
     + \| {\cal F}_{la} \|^2_{\eta,0,{\cal T}} \right) 
     + \| p \|^2_{\eta,1,{\cal T}}  
     + \| \pi \|^2_{\eta,1,{\cal T}}
     + \| q_1 \|^2_{\eta,1,{\cal T}}  \right]\,.
\label{Eq:MixedEstimate1}
\end{equation}
On the other hand, applying the estimate
(\ref{Eq:HigherOrderEstimate}) with $m=2$ to the second order boundary
conditions
(\ref{Eq:SecondOrderBCList4}--\ref{Eq:SecondOrderBCList7}) in the
high frequency limit, we obtain
\begin{eqnarray}
\sum\limits_{a,b \in \{k,m,\bar m\} }
  \left( \eta \| h_{ab} \|^2_{\eta,2,\Omega}
 + \| h_{ab} \|^2_{\eta,2,{\cal T}} \right)
&\leq& C_2
\left[ \sum\limits_{a,b \in \{k,m,\bar m\} } \left( \eta^{-1} 
       \| {\cal F}_{ab} \|^2_{\eta,1,\Omega} 
     + \| {\cal F}_{ab} \|^2_{\eta,0,{\cal T}} \right) \right.
\nonumber\\
     &+& \left. \sum\limits_{a=0}^3 \| h_{la} \|^2_{\eta,2,{\cal T}}
     + \sum\limits_{a=0}^3 \| H_a \|^2_{\eta,1,{\cal T}}
     + \| \psi_0 \|^2_{\eta,0,{\cal T}}  \right]\,.
\label{Eq:MixedEstimate2}
\end{eqnarray}
Combining the two estimates
(\ref{Eq:MixedEstimate1}--\ref{Eq:MixedEstimate2}) we obtain
\begin{eqnarray}
\sum\limits_{a,b=0}^3 \left( \eta \| h_{ab} \|^2_{\eta,2,\Omega}
 + \| h_{ab} \|^2_{\eta,2,{\cal T}} \right)
&\leq& C_3
\left[ \sum\limits_{a,b=0}^3 \left( \eta^{-1} 
       \| {\cal F}_{ab} \|^2_{\eta,1,\Omega} 
     + \| {\cal F}_{ab} \|^2_{\eta,0,{\cal T}} \right) \right.
\nonumber\\
     &+& \left. \sum\limits_{a=0}^3 \| H_a \|^2_{\eta,1,{\cal T}}
     + \| p \|^2_{\eta,1,{\cal T}}  
     + \| \pi \|^2_{\eta,1,{\cal T}}
     + \| q_1 \|^2_{\eta,1,{\cal T}}
     + \| \psi_0 \|^2_{\eta,0,{\cal T}}  \right]\,,
\end{eqnarray}
for a constant $C_3$ which is independent of $\eta > 0$ and
$h_{ab}$. Therefore, we obtain an a priori estimate also in this
case. Notice that we have assumed that $h_{ab}$ and its first two time
derivatives vanish at $t=0$ when deriving this result.

%%%%%%%%%%%%%%%%%%%%%%%%%%%%%%%%%%%%%%%%%%%%%%%%%%%%%%%%%%%%%%
\section{The quality of the boundary conditions}
\label{Sect:Quality}
%%%%%%%%%%%%%%%%%%%%%%%%%%%%%%%%%%%%%%%%%%%%%%%%%%%%%%%%%%%%%%%

In this section, we assess the quality of the boundary conditions
constructed in Sec.~\ref{Sect:OuterBC}. We begin in
Sec.~\ref{Sect:ReflCoeffHighFreq} by computing the reflection
coefficients corresponding to the different boundary conditions in the
high-frequency approximation. Then, in
Sec.~\ref{Sect:ReflCoeffShearBC}, we consider a spherical outer
boundary and compute the reflection coefficient for monochromatic
linearized waves (with not necessarily high frequency) associated with
the new boundary condition (\ref{Eq:FirstOrderRadiativeBC}) on the
shear. Finally, in Sec.~\ref{Sect:NumTests}, we implement the first-
and second order boundary conditions numerically and compare their
performance on a simple test problem.

%%%%%%%%%%%%%%%%%%%%%%%%%%%%%%%%%%%%%%%%%%%%%%%%%%%%%%%%%%%%%%
\subsection{Reflection coefficients in the high-frequency limit}
\label{Sect:ReflCoeffHighFreq}
%%%%%%%%%%%%%%%%%%%%%%%%%%%%%%%%%%%%%%%%%%%%%%%%%%%%%%%%%%%%%%

As shown in the previous section, in the high-frequency limit our
evolution system reduces to a linear constant coefficient problem of
the form (\ref{Eq:WaveProblemEqHigher},\ref{Eq:WaveProblemBCHigher})
on the half-space subject to the harmonic constraint
\begin{equation}
\gz^{ab}\left( \nablaz_a h_{bc} - \frac{1}{2}\nablaz_c h_{ab} \right) = 0\,,
\label{Eq:FrozenCoeffHarmConstr}
\end{equation}
where $\nablaz_a = \partial_a$ since the background metric is
constant. In order to estimate the amount of spurious gravitational
radiation reflected off the boundary, we start with a simplifying
assumption: namely, we assume that the initial data is chosen such
that it is compatible with the harmonic constraint and that the
components $h_{ll}$, $h_{lk}$ and $h_{lm}$ and their time derivatives
are zero. Notice that this is not a restriction on the physics, but
rather a restriction on the choice of coordinates as we show
next. Suppose $h_{ab}$ is an arbitrary solution of
(\ref{Eq:WaveProblemEqHigher}) satisfying the harmonic constraint
(\ref{Eq:FrozenCoeffHarmConstr}). Under an infinitesimal coordinate
transformation $x'^a = x^a + \xi^a$ parametrized by a vector field
$\xi^a$, $h_{ab}$ is mapped to
\begin{equation}
h'_{ab} = h_{ab} + 2\partial_{(a}\xi_{b)} \, .
\label{eq:trans_inf}
\end{equation}
In particular, $h'_{ab}$ still satisfies the harmonic constraint
provided that $\xi_a$ obeys the wave equation
\begin{equation}
0 = \gz^{cd}\nablaz_c\nablaz_d \xi_a 
  = 2\left[ -\nablaz_l\nablaz_k \xi_a + \nablaz_m\nablaz_{\bar{m}} \right]\xi_a
\; .
\label{Eq:WaveEqForXi}
\end{equation}
Requiring $h'_{ll}$, $h'_{lk}$ and $h'_{lm}$ and their time
derivatives to vanish at the initial slice yields the following
conditions:
\begin{eqnarray}
0 = h'_{ll} &=& 
 h_{ll} + \sqrt{2}\,\left[\partial_t \xi_l - (1+\beta)\partial_x\xi_l\right],
\label{eq:coor1}\\
0 = h'_{lk} &=&
 h_{lk} + \frac{1}{\sqrt{2}}\left[
 \partial_t(\xi_l+\xi_k) - (1+\beta)\partial_x\xi_k +(1-\beta)\partial_x\xi_l
\right],
\label{eq:coor2}\\
0 = h'_{lm} &=&
 h_{lm} + \frac{1}{\sqrt{2}}\left[ \partial_t\xi_m - (1+\beta)\partial_x\xi_m
 + (\partial_y + i\,\partial_z)\xi_l\right],
\label{eq:coor3}\\
%%%%%%%%%%%%%%%%%%%%%%%%%%%%%%%%%%%%%%%%%%%%%%%%%%%%
0 = \nablaz_k h'_{ll} &=&
 \nablaz_k h_{ll} + \left(\partial_y^2 + \partial_z^2 \right)\xi_l\; ,
\label{eq:coor4}\\
0 = \nablaz_k(2h'_{lk} - h'_{ll}) &=&
 \nablaz_k(2h_{lk} - h_{ll}) + 2\sqrt{2}\partial_x\nablaz_k\xi_l
 + \left(\partial_y^2 + \partial_z^2 \right)\xi_k\; ,
\label{eq:coor5}\\
0 = 2\nablaz_k h'_{lm} &=& 2\nablaz_k h_{lm} + 2\nablaz_m\nablaz_k\xi_k 
 + \left(\partial_y^2 + \partial_z^2 \right)\xi_m\; ,
\label{eq:coor6}
\end{eqnarray}
where we have used the tetrad fields (\ref{eq:tetrad}) and
Eq. (\ref{Eq:WaveEqForXi}). Equations
(\ref{eq:coor1})--(\ref{eq:coor3}) yield elliptic equations on each
$x=const.$ surface for $\xi_l$, $\xi_k$ and $\xi_m$ and can be solved
provided appropriate fall-off conditions at $y^2 + z^2 \to \infty$ are
specified. Once these equations are solved,
Eqs. (\ref{eq:coor1}--\ref{eq:coor3}) can be solved for
$\partial_t\xi_l$, $\partial_t\xi_k$ and $\partial_t\xi_m$.
Therefore, it is always possible to choose the gauge such that the
harmonic constraint is satisfied and such that initially, $h'_{ll}$,
$h'_{lk}$ and $h'_{lm}$ and their time derivatives vanish.

The evolution equations for $h_{ab}$, the boundary conditions
(\ref{Eq:FirstOrderBCList1}--\ref{Eq:FirstOrderBCList3})
or
(\ref{Eq:HigherOrderGaugeBC1}--\ref{Eq:HigherOrderGaugeBC3})
which, in the high-frequency limit, reduce to
\begin{displaymath}
\left[ \partial_t - (1 + \beta)\partial_x \right]^{L+1} u \hateq 0\,,
\qquad u = h_{ll}, h_{lk}, h_{lm}\; ,\qquad L \geq 0\,,
\end{displaymath}
and the well posedness result derived in the previous section imply
that $h_{ll} = h_{lk} = h_{lm} = 0$ everywhere and at all times. In
this gauge, the harmonic constraint (\ref{Eq:FrozenCoeffHarmConstr})
yields
\begin{displaymath}
\nablaz_l h_{m\bar{m}} = 0, \qquad
\nablaz_l h_{kk} = -\nablaz_k h_{m\bar{m}} + 2\nablaz_{(m} h_{\bar{m})k}\; ,
\qquad
\nablaz_l h_{km} = \nablaz_{\bar{m}} h_{mm}\; .
\end{displaymath}
The first condition, together with the wave equation
$-\nablaz_k\nablaz_l h_{m\bar{m}} + \nablaz_{(m}\nablaz_{\bar{m})}
h_{m\bar{m}} = 0$ implies that $h_{m\bar{m}} = 0$ provided appropriate
fall-off conditions at $y^2 + z^2 \to \infty$ are specified. Knowing
$h_{mm}$, the third equation can then be integrated along $l^a$ to
obtain $h_{km}$. Since $l^a$ is outgoing at the boundary, the initial
data for $h_{km}$ completely determines the solution. Once $h_{km}$ is
known, the second equation can be integrated in order to obtain
$h_{kk}$.

Therefore, in the gauge where $l^a h_{ab} = 0$, the entire dynamics is
governed by the evolution system
\begin{eqnarray}
&&\left[ -\partial_t^2 + 2\beta\partial_t\partial_x + (1-\beta^2)\partial_x^2
 + \partial_y^2 + \partial_z^2 \right] h_{mm} = 0\,,
\label{Eq:WaveEqRefl}\\
&&\left[ \partial_t - (1 + \beta)\partial_x \right]^{L+1} h_{mm} \hateq 0\,,
\label{Eq:WaveBCRefl}
\end{eqnarray}
where $L+1=1,2,3,...$ is the order of the boundary condition
considered. In order to quantify the amount of spurious reflections,
we consider a monochromatic plane wave with frequency $\omega > 0$ and
wave vector $(p_j) = q(-1,\tan(\theta),0)$ with $q > 0$ and $\theta
\in (-\pi/2,\pi/2)$ the angle of incidence, which is reflected off the
boundary $x=0$. Therefore, the solution has the form
\begin{displaymath}
h_{mm} = e^{i\,(\omega\,t - p_j x^j)} 
       + \gamma\, e^{i\,(\omega\,t - \hat{p}_j x^j)}\, ,\qquad
(x^j) = (x,y,z) \in \Sigma\,,
\end{displaymath}
where $(\hat{p}_j) = q(1,\tan(\theta),0)$ and $\gamma$ is an amplitude
reflection coefficient. Introducing this ansatz into the wave equation
(\ref{Eq:WaveEqRefl}) and the boundary condition (\ref{Eq:WaveBCRefl})
yields the dispersion relation
\begin{displaymath}
\omega = q\left[\,\beta + \sqrt{1 + \tan^2(\theta)}\,\right]\,,
\end{displaymath}
and the reflection coefficient
\begin{equation}
\gamma = -\left[ \frac{1 - \cos\theta}{1 + (1+2\beta)\cos\theta} \right]^{L+1}\,.
\label{Eq:ReflCoeff1}
\end{equation}
In particular, $\gamma=0$ for normal incidence and $\gamma\to -1$ for
modes propagating tangential to the boundary ($\theta\to \pm\pi/2$).
For modes with fixed incidence angle $-\pi/2 < \theta < \pi/2$ the
square bracket is nonnegative and strictly smaller than one which
shows that fewer and fewer reflections are present if $L$ is
increased. For $\beta=0$ and $L=0,1$, the expression for $\gamma$
given in (\ref{Eq:ReflCoeff1}) agrees with the coefficients obtained
in Sec.~1.B of Ref. \cite{Engquist77}.

Finally, we notice that the reflection coefficient $\gamma$ depends
neither on the frequency nor on the wavelength. As we will see in the
next subsection, this is an artifact of the high-frequency
approximation. Also, we would like to stress that the result
(\ref{Eq:ReflCoeff1}) relies on the gauge choice $l^a h_{ab} = 0$
which we adopted here; it might change for other coordinate choices.

%%%%%%%%%%%%%%%%%%%%%%%%%%%%%%%%%%%%%%%%%%%%%%%%%%%%%%%%%%%%%%
\subsection{Reflection coefficients for the shear boundary condition}
\label{Sect:ReflCoeffShearBC}
%%%%%%%%%%%%%%%%%%%%%%%%%%%%%%%%%%%%%%%%%%%%%%%%%%%%%%%%%%%%%%

Next, we generalize the above analysis by relaxing the high frequency
assumption. Instead, we assume that close to the boundary surface,
spacetime can be written as the Schwarzschild metric of mass $M$
(where $M$ denotes the ADM mass of the system) plus a small
perturbation thereof. Assuming that the outer boundary is an
approximate sphere with areal radius $R \gg M$ and considering
monochromatic gravitational radiation characterized by a wave number
$k \gg M^{-1}$, it has been shown in \cite{Buchman:2006xf} that the
freezing-$\Psi_0$ boundary condition (\ref{Eq:SecondOrderBCList4})
yields a reflection coefficient which is of the order of $(k R)^{-4}$
for quadrupolar gravitational radiation. Reflection coefficients for
the higher order boundary conditions (\ref{Eq:HigherOrderPsi0}) were
also computed in \cite{Buchman:2006xf,Buchman:2007pj} with the result $(k
R)^{-2(L+1)}$ for waves with multipole moment $\ell > L$.

Here, we want to compute the reflection coefficient for our shear
boundary condition (\ref{Eq:FirstOrderRadiativeBC}). As shown at the
end of Sec.~\ref{Sect:OuterBC_third} it can be considered as the
$L=0$ member of the hierarchy of absorbing boundary conditions under
certain circumstances such as the frozen coefficient limit in the
gauge $l^a h_{ab} = 0$. In view of this, one could hope for a
reflection coefficient of the order $(k R)^{-2}$ for quadrupolar
radiation. Unfortunately, as we show now the reflection coefficient
only scales as $(kR)^{-1}$ for large $k R$.

In order to analyze this, consider odd-parity linear perturbations of
the Schwarzschild spacetime $(M,\gz)$. Hence, the manifold has the
form $M = \tilde{M} \times S^2$ and the background metric the form
\begin{equation}
\gz_{ab} = \tilde{g}_{ij}\,dx^i\,dx^j + r^2\,\hat{g}_{AB}\,dx^A\,dx^B\,,
\label{Eq:sphere}
\end{equation}
where $\tilde{g}_{ij}$ denotes a pseudo-Riemannian metric on the
two-dimensional orbit manifold $\tilde{M}$, $r$ is the areal radius
and $\hat{g}_{AB}$ is the standard metric on $S^2$. Metric
perturbations of such a background have the following form
\begin{displaymath}
\delta g_{ij} = L_{ij}\; ,\qquad
\delta g_{Aj} = Q_{Aj}\; ,\qquad
\delta g_{AB} = r^2 K_{AB}\; ,
\end{displaymath}
where the quantities $L_{ij}$, $Q_{Aj}$ and $K_{AB}$ depend on the
coordinates $x^i$ and $x^A$. Using this notation, one finds that 
to linear order in the perturbation, the shear (\ref{Eq:shear})
associated with a $t=const$, $r=const$ surface is given by
\begin{eqnarray}
&&\delta\sigma_{ij} = 0\,, \nonumber\\
&&\delta\sigma_{Aj} = 0\,, \\
&&\delta\sigma_{AB} = \frac{1}{2}\,l^j\,
\left[r^2\,\tilde{\nabla}_j\,\hat{K}_{AB} - 2\,\hat{\nabla}_{(A}Q_{B)j} +
\hat{g}_{AB}\,\hat{g}^{CD}\,\hat{\nabla}_C\,Q_{Dj}\right]\, ,
\nonumber
\label{Eq:shearpertu}
\end{eqnarray}
where here $\tilde{\nabla}$ and $\hat{\nabla}$ refer to the covariant
derivative associated with $\tilde{g}_{ij}$ and $\hat{g}_{AB}$,
respectively, and $\hat{K}_{AB} = K_{AB} - \hat{g}_{AB}\,\hat{g}^{CD}
K_{CD}/2$ is the trace-free part of $K_{AB}$. Using the transformation
properties of the fields $L_{ij}$, $Q_{Aj}$ and $K_{AB}$ under
infinitesimal coordinate transformations (see for instance
Ref.~\cite{Buchman:2007pj}) one can check that $\delta\sigma_{AB}$ is
invariant with respect to odd-parity coordinate transformations (but
not under transformations with even parity). For this reason, in the
following, we restrict our attention to odd-parity perturbations since
in this case the shear boundary condition
(\ref{Eq:FirstOrderRadiativeBC}) has a gauge-invariant interpretation.

\begin{figure}
\epsfig{file=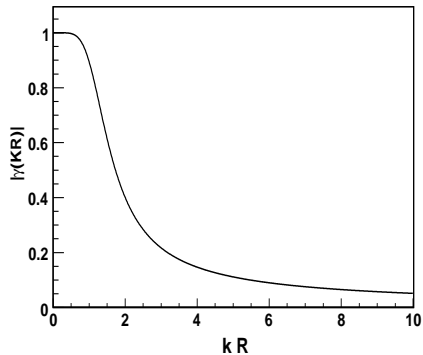,width=60mm,height=50mm}
\caption{Reflection coefficient $|\gamma(kR)|$ as a function of $kR$
for the shear boundary condition (\ref{Eq:FirstOrderRadiativeBC}) for
weak, monochromatic quadrupolar waves with wave number $k$ and odd
parity. The reflection coefficient is of order unity for small $kR$
and decays as $(kR)^{-1}$ for large $k R$.}
\label{fig:reflection_c}
\end{figure}

Perturbations with odd parity and fixed angular momentum numbers $\ell,
m$ are parametrized by a scalar field $\kappa$ and a one-form $h = h_a
\,dx^a$ on $\tilde{M}$ according to
\begin{displaymath}
L_{ij} = 0\,, \qquad
Q_{Aj} = h_j\,S_A\; , \qquad
K_{AB} = 2\kappa\,\hat{\nabla}_{(A}S_{B)}\; ,
\end{displaymath}
where $S_A = \hat{\varepsilon}_A{}^B\hat{\nabla}_B\,Y$ with
$\hat{\varepsilon}_{AB}$ the natural volume element on $S^2$ and $Y
\equiv Y^{\ell m}$ the standard spherical harmonics. With this
notation we obtain
\begin{displaymath}
\delta\sigma_{AB} = -l^j h^{(inv)}_j \,\hat{\nabla}_{(A}S_{B)}\; ,
\end{displaymath}
where $h^{(inv)}_j$ is the gauge-invariant one-form \cite{Gerlach79}
\begin{displaymath}
h^{(inv)}_j = h_j - r^2\,\tilde{\nabla}_j\left(\frac{\kappa}{r^2}\right)\,.
\end{displaymath}
In terms of the gauge-invariant scalar $\Phi$ obeying the Regge-Wheeler
equation \cite{Regge57,Sarbach:1999gi,Sarbach:2001qq}
\begin{equation}
\left[ -\tilde{g}^{ij}\tilde{\nabla}_i\tilde{\nabla}_j 
 + \frac{\ell(\ell+1)}{r^2} - \frac{6M}{r^3} \right]\Phi = 0\,,
\label{Eq:RW}
\end{equation}
this gauge-invariant one-form can be computed according to
$h^{(inv)}_j = \tilde{\varepsilon}_{ij}\,\tilde{\nabla}^i(r\,\Phi)$,
where $\tilde{\varepsilon}_{ij}$ is the induced volume element on
$\tilde{M}$ \cite{Gerlach79,Sarbach:1999gi,Sarbach:2001qq}. Therefore,
the shear boundary condition (\ref{Eq:FirstOrderRadiativeBC}) implies
the following boundary condition for the Regge-Wheeler equation
governing the dynamics of gravitational perturbations with odd parity,
\begin{equation}
l^j\tilde{\nabla}_j(r\,\Phi) \hateq 0\,.
\label{Eq:shear_invar}
\end{equation}
For the following, we assume that the coordinates $x^i = (t,r)$ are
such that
\begin{displaymath}
\tilde{g}_{ij} dx^i dx^j = -dt^2 + dr^2 + O\left( \frac{2M}{R} \right)\,, 
\qquad r \approx R\,.
\end{displaymath}
In order to quantify the amount of spurious reflections generated by
the boundary conditions (\ref{Eq:shear_invar}), we impose these
conditions at finite radius $r = R < \infty$ and, following
\cite{Buchman:2006xf}, consider monochromatic quadrupolar waves of the form
\begin{equation}
\Phi(t,r) = a_2^\dagger\,a_1^\dagger\left(e^{i\,k\,(r-t)} 
               + \gamma\,e^{-i\,k\,(r+t)} \right) 
 + O\left( \frac{2M}{R} \right)\,,
\label{Eq:reflection}
\end{equation}
where $a_2^\dagger = -\partial_r + 2/r$, $a_1^\dagger = -\partial_r +
1/r$, $k > 0$ is a given wave number and $\gamma$ is the amplitude
reflection coefficient. Introducing (\ref{Eq:reflection}) into the
boundary condition (\ref{Eq:shear_invar}) yields (neglecting the
$2M/R$ correction terms)
\begin{displaymath}
-e^{2ikR}\,\left[3+(kR)^2\right] + \gamma\,\left(kR-i\right)
\left[2i(kR)^2+3\,kR-3\,i\right] = 0\,.
\end{displaymath}
Solving for $\gamma$, the amount of reflection is given by
\begin{equation}
  \label{Eq:SchwReflCoeff}
|\gamma(kR)| = \left[{1
 + \frac{4\,(k\,R)^6}{\left[3+(k\,R)^2\right]^2}}\right]^{-1/2}\,.
\end{equation}
The reflection coefficient $|\gamma(kR)|$ is shown in
Fig.~(\ref{fig:reflection_c}). It can be seen from
Eq.~(\ref{Eq:SchwReflCoeff}) that the coefficient decays as
$(kR)^{-1}$ for large $kR$. This is slower than the $(kR)^{-2}$ decay
we had hoped for. Therefore, it is worthwhile investing the effort to
implement the second order boundary condition,
Eq.~(\ref{Eq:SecondOrderBCList4}), which specifies $\Psi_0$ and yields
a reflection coefficient that decays as $(k R)^{-4}$ when $\Psi_0$ is
frozen to its initial value.

%%%%%%%%%%%%%%%%%%%%%%%%%%%%%%%%%%%%%%%%%%%%%%%%%%%%%%%%%%%%%%
\subsection{Numerical tests}
\label{Sect:NumTests}
%%%%%%%%%%%%%%%%%%%%%%%%%%%%%%%%%%%%%%%%%%%%%%%%%%%%%%%%%%%%%%

An ideal boundary condition would produce a solution that is identical
(within the computational domain) to the corresponding solution on an
unbounded domain. This principle was used in \cite{Rinne:2007ui} to
assess the numerical performance of various boundary conditions.
First, a \emph{reference solution} is computed on a very large
computational domain. Next, the domain is truncated at a smaller
distance where the boundary conditions are imposed.  The reference
domain is chosen large enough such that its boundary remains out of
causal contact with the smaller domain for as long as we evolve.
Finally, the solution on the smaller domain is compared with the
reference solution, measuring the spurious reflections and constraint
violations caused by the boundary conditions.

Here we use the same test problem as in \cite{Rinne:2007ui}. The
initial data are taken to be a Schwarzschild black hole of mass $M$ in
Kerr-Schild coordinates with an outgoing odd-parity quadrupolar
gravitational wave perturbation (satisfying the full nonlinear
constraint equations).  The perturbation is centered about a radius
$r_0 = 5 M$ initially and its dominant wavelength is $\lambda \approx
4 M$.

These initial data are evolved on a spherical shell extending from $r
= 1.9 M$ (just inside the horizon; no boundary conditions are needed
here) out to $R = 961.9 M$ for the reference solution and to $R = 41.9
M$ for the truncated domain.  The gauge source functions $H_a$ are
chosen initially such that the time derivatives of the lapse and shift
vanish, see
Eqs. (\ref{Eq:HarmConstrAlphaDot},\ref{Eq:HarmConstrBetaDot}). This
value of $H_a$ is then frozen in time.

A first order formulation (in both space and time) of the generalized
harmonic Einstein equations is used as described in \cite{Lindblom06}.
Our numerical implementation employs the Caltech-Cornell Spectral
Einstein Code (SpEC), which is based on a pseudospectral collocation
method.  We refer the reader to appendix A of \cite{Rinne:2007ui} for
details on the numerical method, the test problem, and the various
diagnostic quantities discussed below.

Four different sets of boundary conditions are compared,
\begin{enumerate}
  \item 
    the first order conditions 
    (\ref{Eq:FirstOrderBCList1})--(\ref{Eq:FirstOrderBCList7}),
    which include the vanishing shear condition (\ref{Eq:FirstOrderBCList4}),
  \item
    the original Kreiss-Winicour \cite{Kreiss:2006mi} boundary conditions, 
    which replace Eq.~(\ref{Eq:FirstOrderBCList4}) with
    \begin{equation}
      D_{lmm} \hateq q_2'\,,
    \end{equation}
    and are otherwise identical to the previous set,
  \item
    the second order constraint-preserving boundary conditions 
    with $\Psi_0$ freezing, 
    Eqs.~(\ref{Eq:SecondOrderBCList1})--(\ref{Eq:SecondOrderBCList7}), 
  \item
    the same as the previous set but with the first order gauge boundary 
    conditions (\ref{Eq:FirstOrderBCList1})--(\ref{Eq:FirstOrderBCList3})
    instead of the second order ones 
    (\ref{Eq:SecondOrderBCList1})--(\ref{Eq:SecondOrderBCList3});
    these are the boundary conditions used in 
    \cite{Lindblom06,Rinne:2006vv,Rinne:2007ui}.
\end{enumerate}
Our implementation of the gauge boundary conditions differs from
Eqs.~(\ref{Eq:FirstOrderBCList1})--(\ref{Eq:FirstOrderBCList3})
or~(\ref{Eq:SecondOrderBCList1})--(\ref{Eq:SecondOrderBCList3}) by
terms of lower derivative order, which were found experimentally to
slightly reduce reflections from the outer boundary in the components
$l^a h_{ab}$ of the metric. Such non-principal terms do not affect
the well posedness results of Sec.~\ref{Sect:WellPosedness}.

\begin{figure}
  \plot{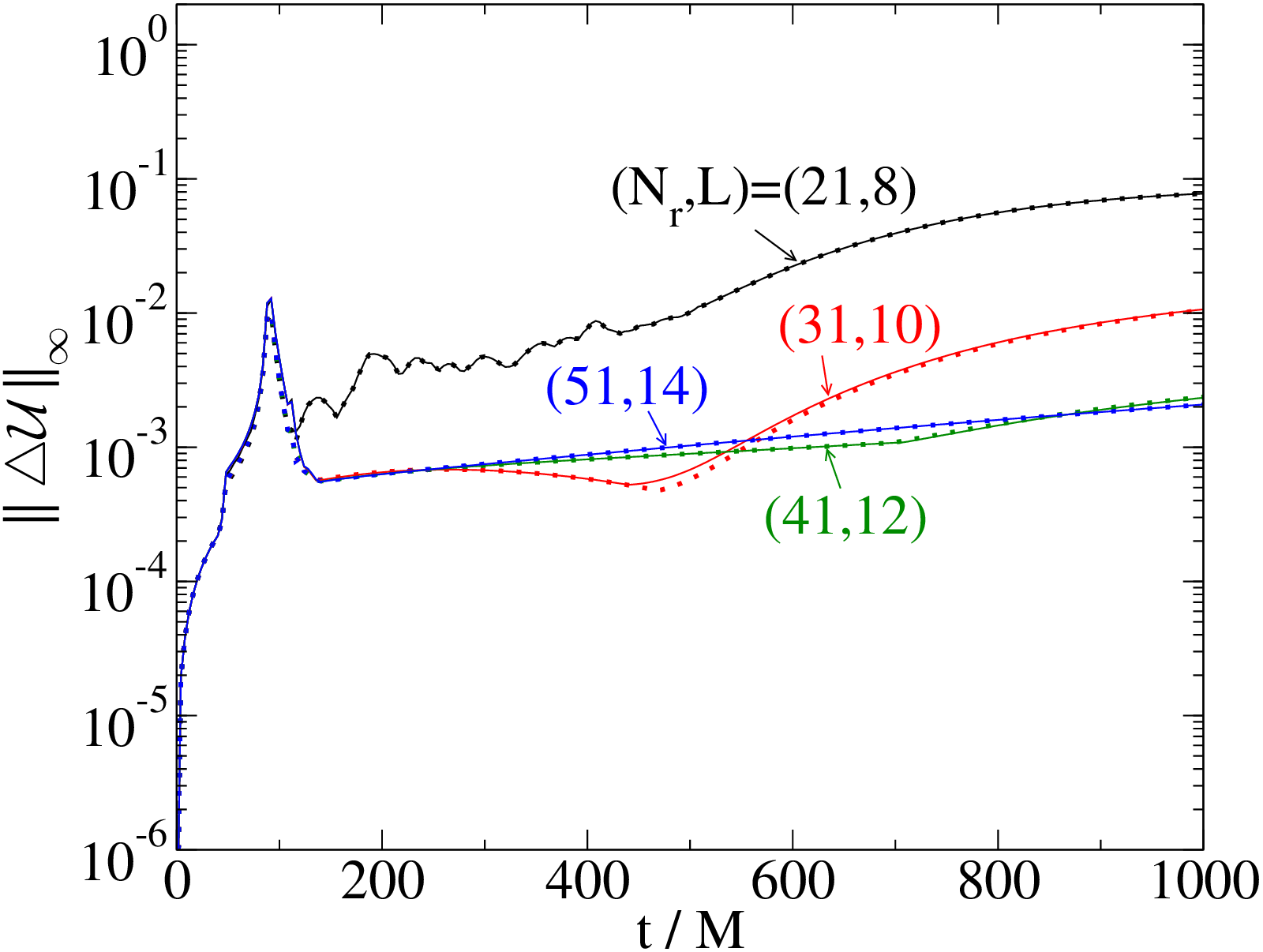}
  \plot{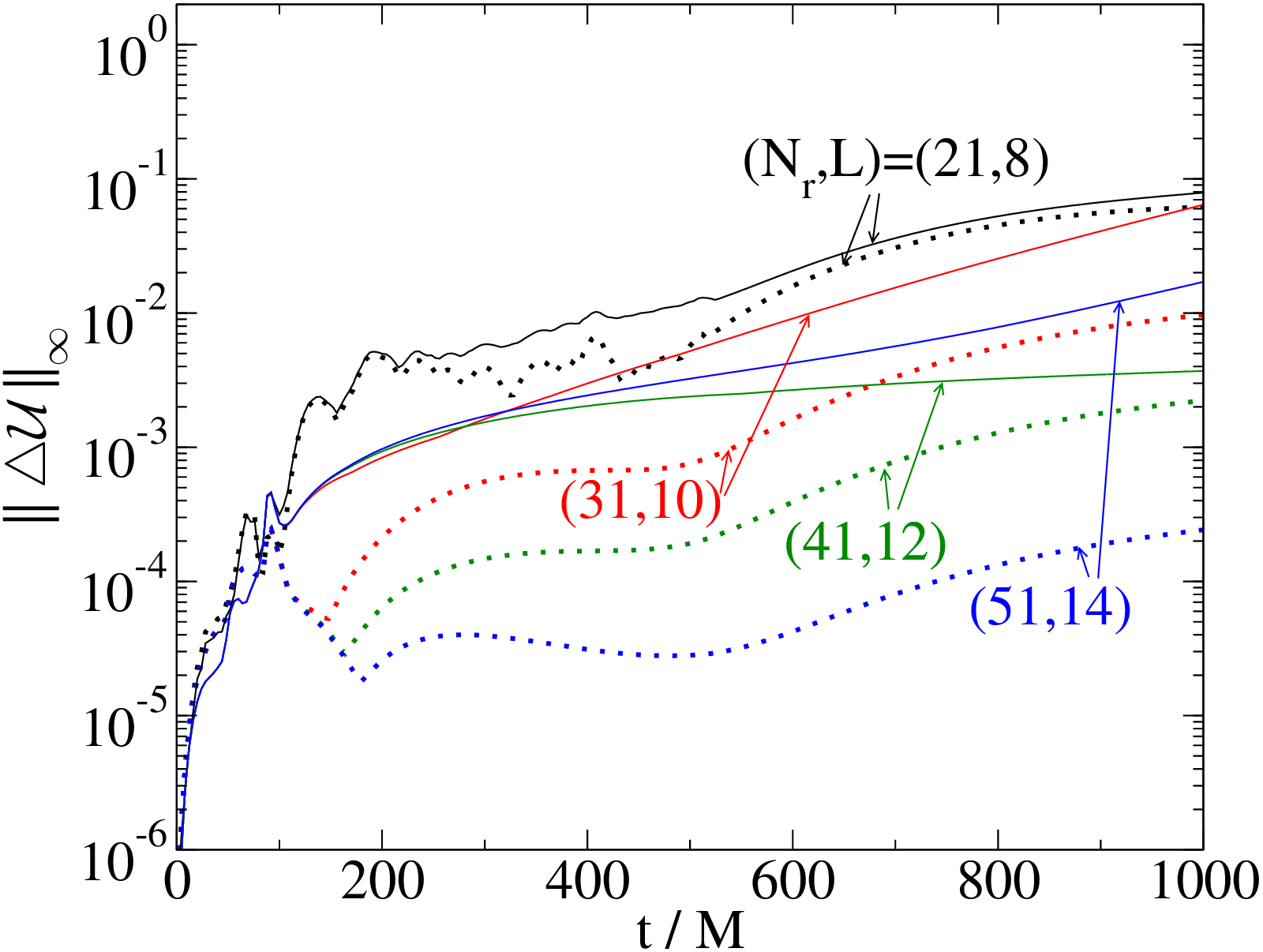}
  \caption{\label{fig:Differences} 
    Difference $\Delta \mathcal{U}$ with respect to the reference solution 
    for four different resolutions $(N_r, L)$.
    Left: 1.~vanishing shear (solid) vs.~2.~Kreiss-Winicour (dotted) 
    boundary conditions,
    right: 3.~second order boundary conditions (solid) vs.~4.~second order
    boundary conditions with first order gauge boundary conditions (dotted).
}
\end{figure}  

Fig.~\ref{fig:Differences} shows the $L^\infty$ norm of the difference
$\Delta \mathcal{U}$ of the solution on the truncated domain with
respect to the reference solution as a function of time.  This
quantity is obtained by taking a tensor norm of the differences in the
metric and its first derivatives at each point \cite{Rinne:2007ui}.
We normalize $\Delta \mathcal{U}$ by the analogous difference of the
perturbed initial data with respect to the unperturbed data. The
results for both versions of the first order boundary conditions are
very similar.  A first peak arises when the reflection from the outer
boundary reaches the center, where its amplitude assumes its maximum
because of the spherical geometry. For the second order boundary
conditions, the peak is smaller by about two orders of magnitude.  For
the second order boundary conditions with first order gauge boundary
conditions, $\Delta \mathcal{U}$ appears to converge away even for the
higher resolutions at late times, unlike for the first order
conditions. Unfortunately, this is not the case for the second order
gauge boundary conditions.  For those, $\Delta \mathcal{U}$ grows at
late times at a rate that does not appear to depend on resolution in a
monotonous way.  A closer look at the data indicates that this growth
only affects the $L = 1, 2$ spherical harmonic basis functions.  We
suspect that this is a numerical problem related to spectral filtering
(cf. \cite{Lindblom05}); so far we have not been able to cure it.
Note that $\Delta \mathcal{U}$ is a gauge-dependent quantity because
the difference norm includes the entire spacetime metric. In fact, as
we shall see below, inspection of the errors in the constraints and in
the Newman-Penrose scalar $\Psi_4$ (which can be viewed as an
approximation to the outgoing gravitational radiation) suggests that
the blow-up is a pure gauge effect.

\begin{figure}
  \plot{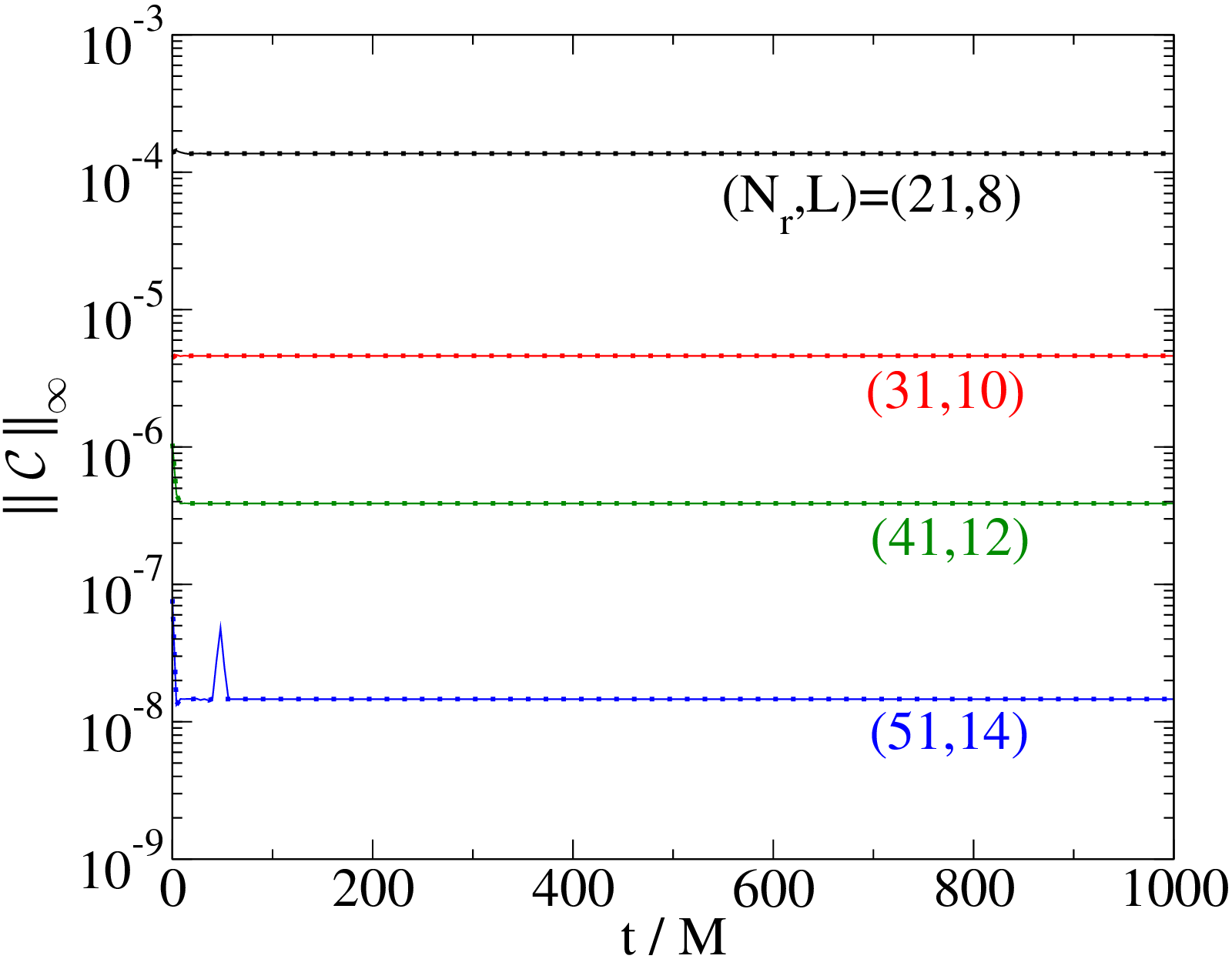}
  \plot{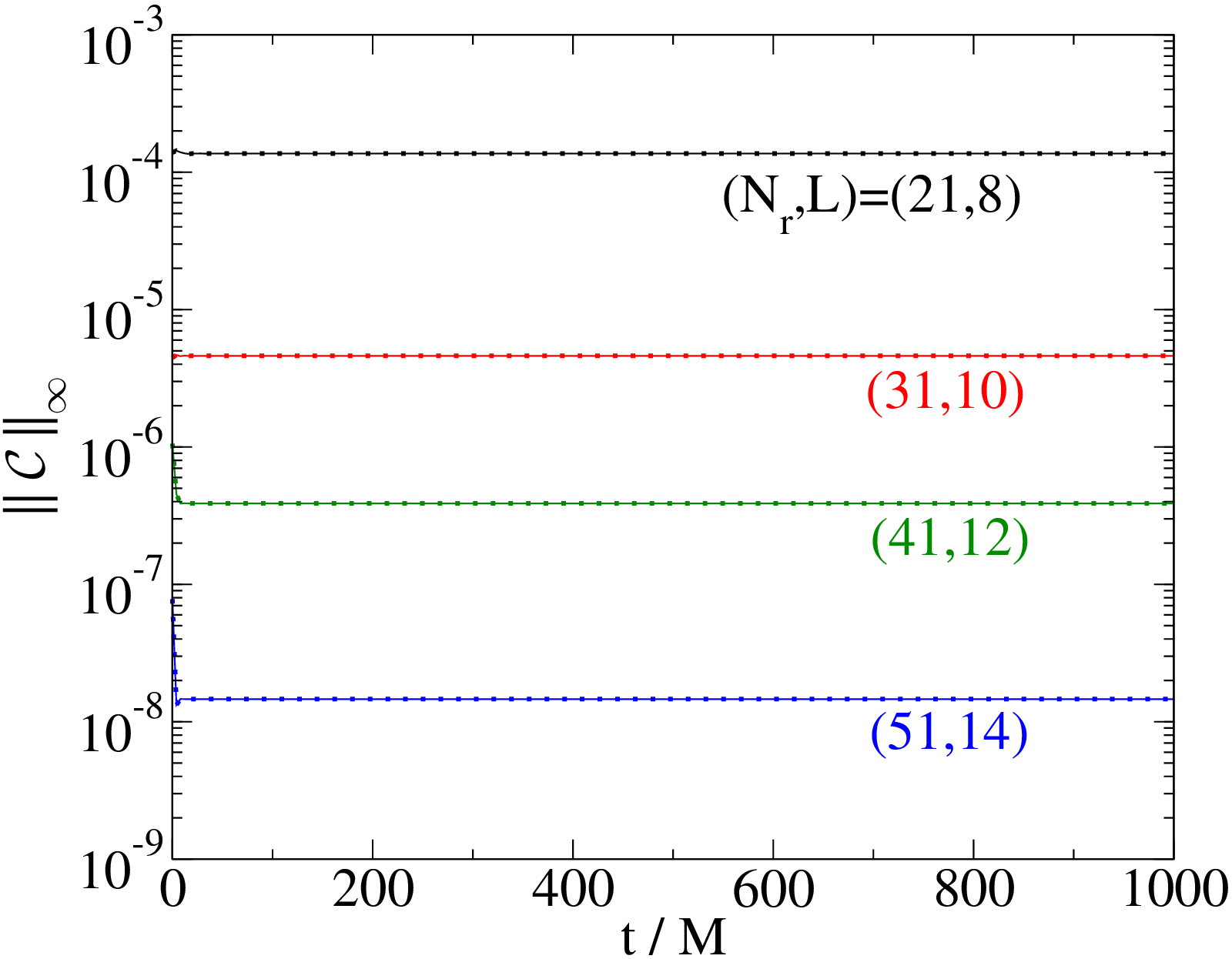}
  \caption{\label{fig:Constraints} 
    Constraint violations $\mathcal{C}$ 
    for four different resolutions $(N_r, L)$. 
    Left: 1.~vanishing shear (solid) vs.~2.~Kreiss-Winicour (dotted) 
    boundary conditions,
    right: 3.~second order boundary conditions (solid) vs.~4.~second order
    boundary conditions with first order gauge boundary conditions (dotted).
}
\end{figure}  

The violations of the constraints are shown in
Fig.~\ref{fig:Constraints}. The quantity $\mathcal{C}$ is a tensor norm
including the harmonic constraints (\ref{Eq:HarmConstr}) as well as
the additional constraints arising from the first order reduction of
\cite{Lindblom06}. We normalize $\mathcal{C}$ by the second
derivatives of the metric so that $\mathcal{C} \sim 1$ means that the
constraints are not satisfied at all.  The constraint violations
converge away with increasing resolution for all the boundary
conditions. This is what we expect because all the boundary conditions
we considered are constraint-preserving.

\begin{figure}
  \plot{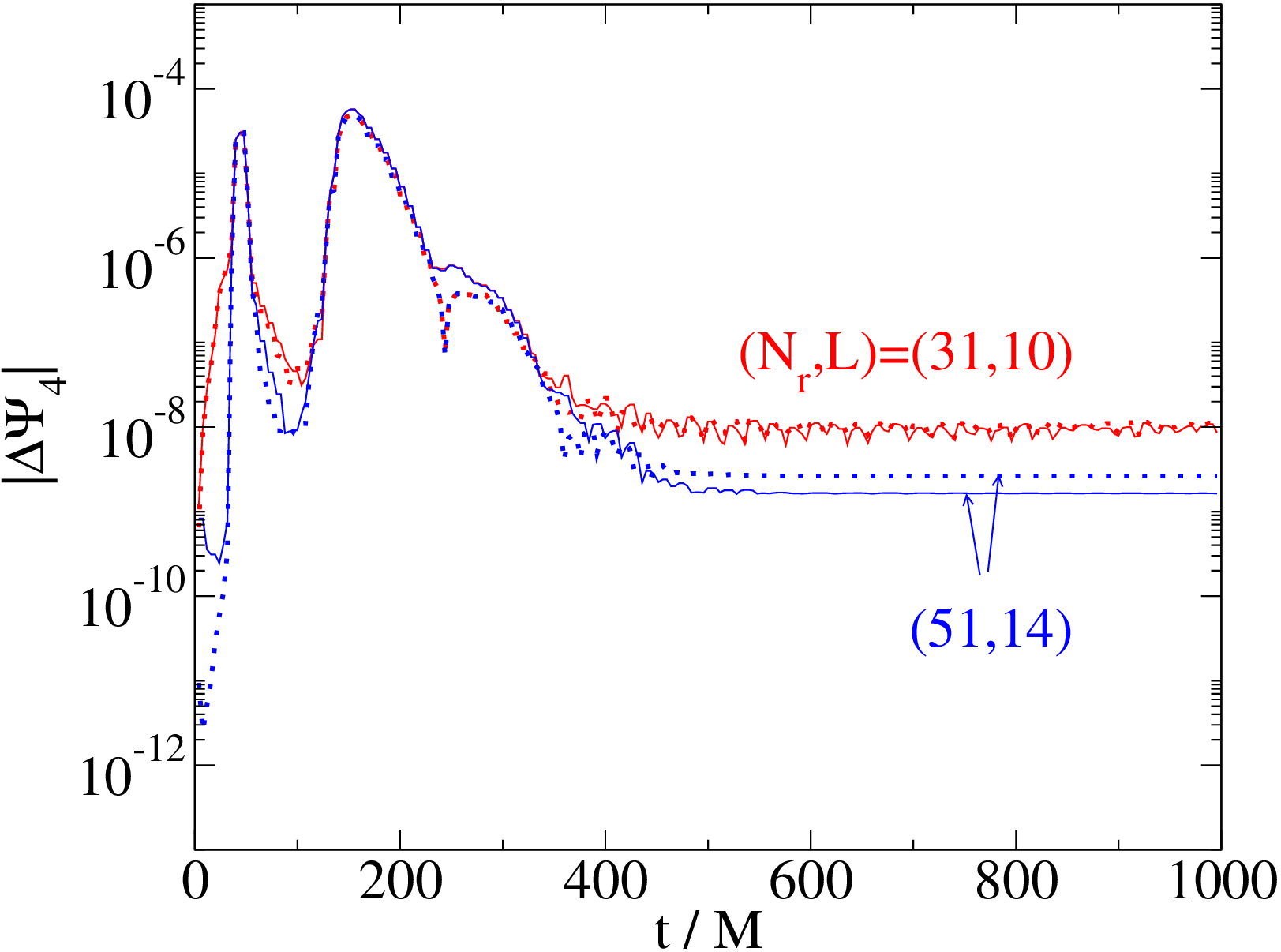}
  \plot{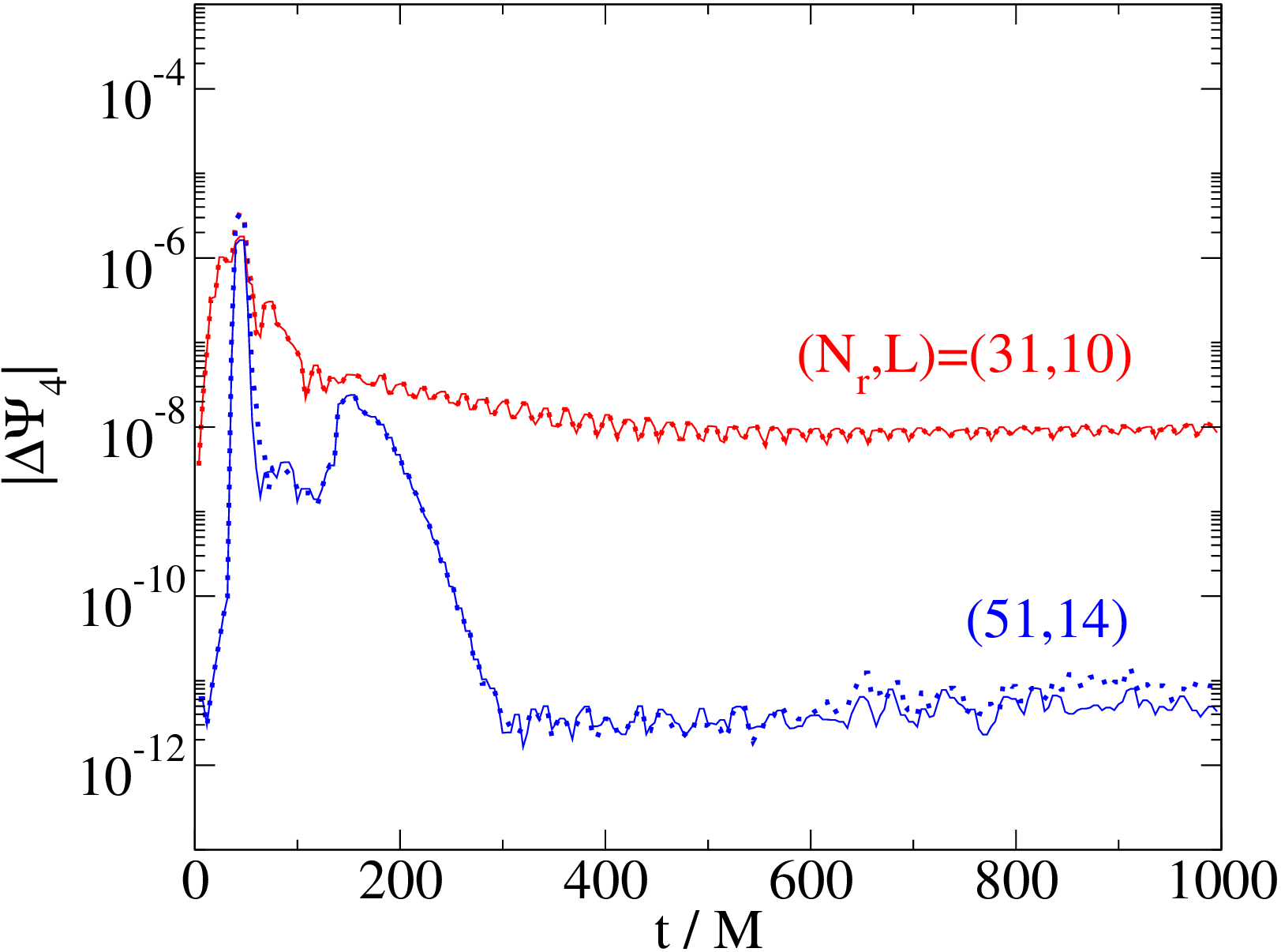}
  \caption{\label{fig:DeltaPsi4} 
    Difference of $\Psi_4$ with respect to the reference solution 
    for two different resolutions.
    Left: 1.~vanishing shear (solid) vs.~2.~Kreiss-Winicour (dotted) 
    boundary conditions,
    right: 3.~second order boundary conditions (solid) vs.~4.~second order
    boundary conditions with first order gauge boundary conditions (dotted).
}
\end{figure}  

One of the main objectives of numerical relativity is the computation
of the gravitational radiation emitted by a compact source.  Hence it
is important to evaluate how the boundary conditions affect the
accuracy of the extracted waveform.  To this end, we compute the
Newman-Penrose scalar $\Psi_4$ on an extraction sphere close to the
outer boundary (at $R_\mathrm{ex} = 40 M$)\footnote{We decompose the
  Newman-Penrose scalars with respect to spin-weighted spherical
  harmonics on the extraction sphere and only display the (by far) 
  dominant mode \cite{Rinne:2007ui}.}. 
The tetrad we use agrees
with the one given in Eqs.~(\ref{Eq:NPTetrad1})--(\ref{Eq:NPTetrad2})
when evaluated for the background spacetime (see \cite{Rinne:2007ui}
for details). Strictly speaking, $\Psi_4$ only has a gauge-invariant
meaning in the limit as future null infinity is approached but since
our computational domain does not extend to infinity we can only
evaluate $\Psi_4$ at a finite radius. However, $\Psi_4$ is
gauge-invariant with respect to infinitesimal coordinate
transformations and tetrad rotations on a Schwarzschild background, so
errors in $\Psi_4$ due to gauge ambiguities should be very small.
Fig.~\ref{fig:DeltaPsi4} shows the difference of $\Psi_4$ with respect
to the same quantity obtained from the reference solution at the same
location.  We normalize $|\Delta \Psi_4|$ by the maximum in time of
$|\Psi_4|$ at the extraction radius.  Again, both versions of the first
order boundary conditions show very similar numerical performance.
Clearly visible is a first peak arising when the outgoing wave passes
through the extraction sphere.  Some of it is reflected off the
boundary and excites the black hole, which then emits quasinormal mode
radiation of exponentially decaying amplitude--a feature also visible
in Fig.~\ref{fig:DeltaPsi4}.  The reflections are much smaller for
(both versions of) the second order boundary conditions (about an
order of magnitude at the first peak and two--three orders of
magnitude later on).  Unlike for the first order conditions, their
$|\Delta \Psi_4|$ decreases with increasing resolution, at least at
late times.

\begin{figure}
  \plot{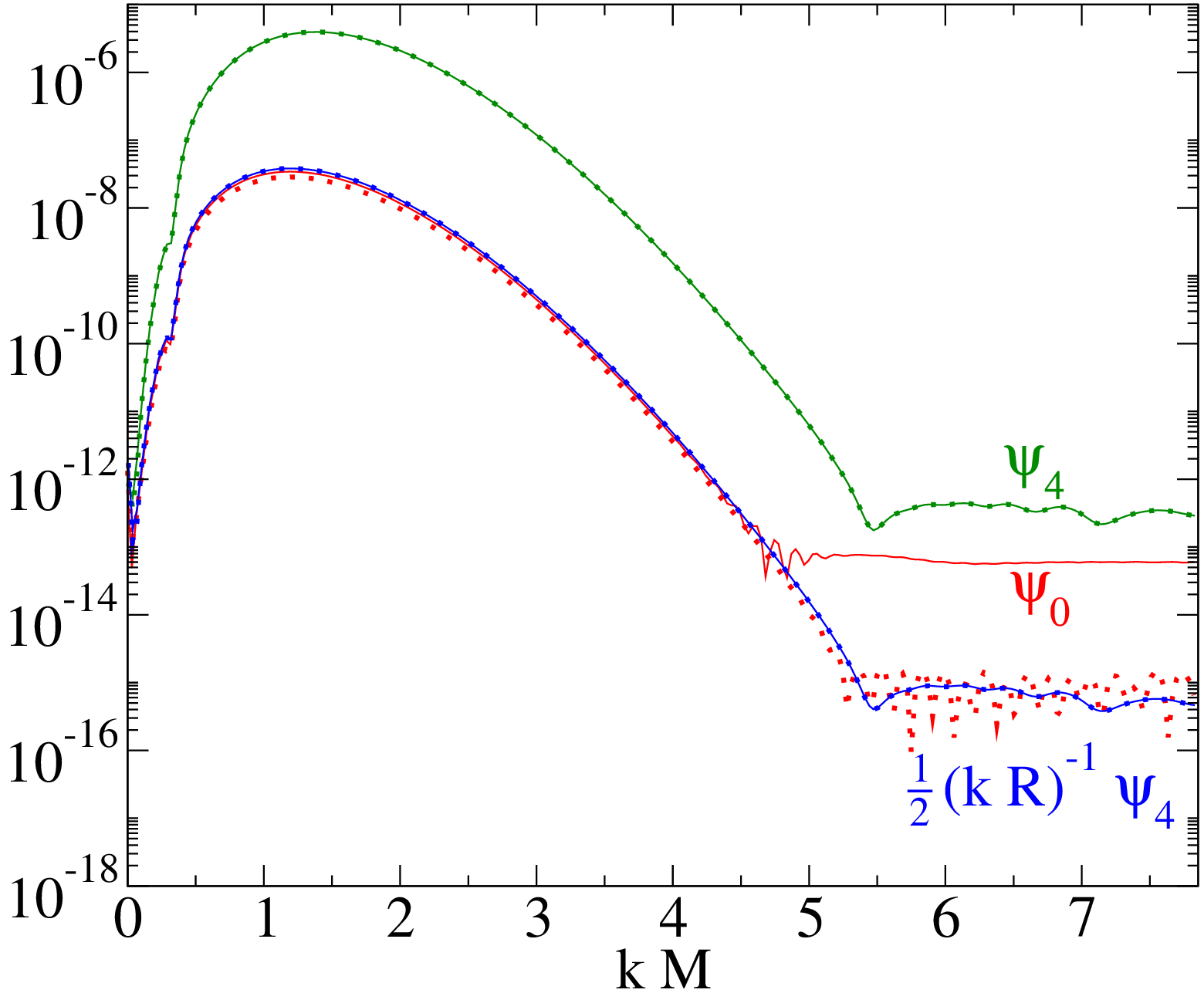}
  \plot{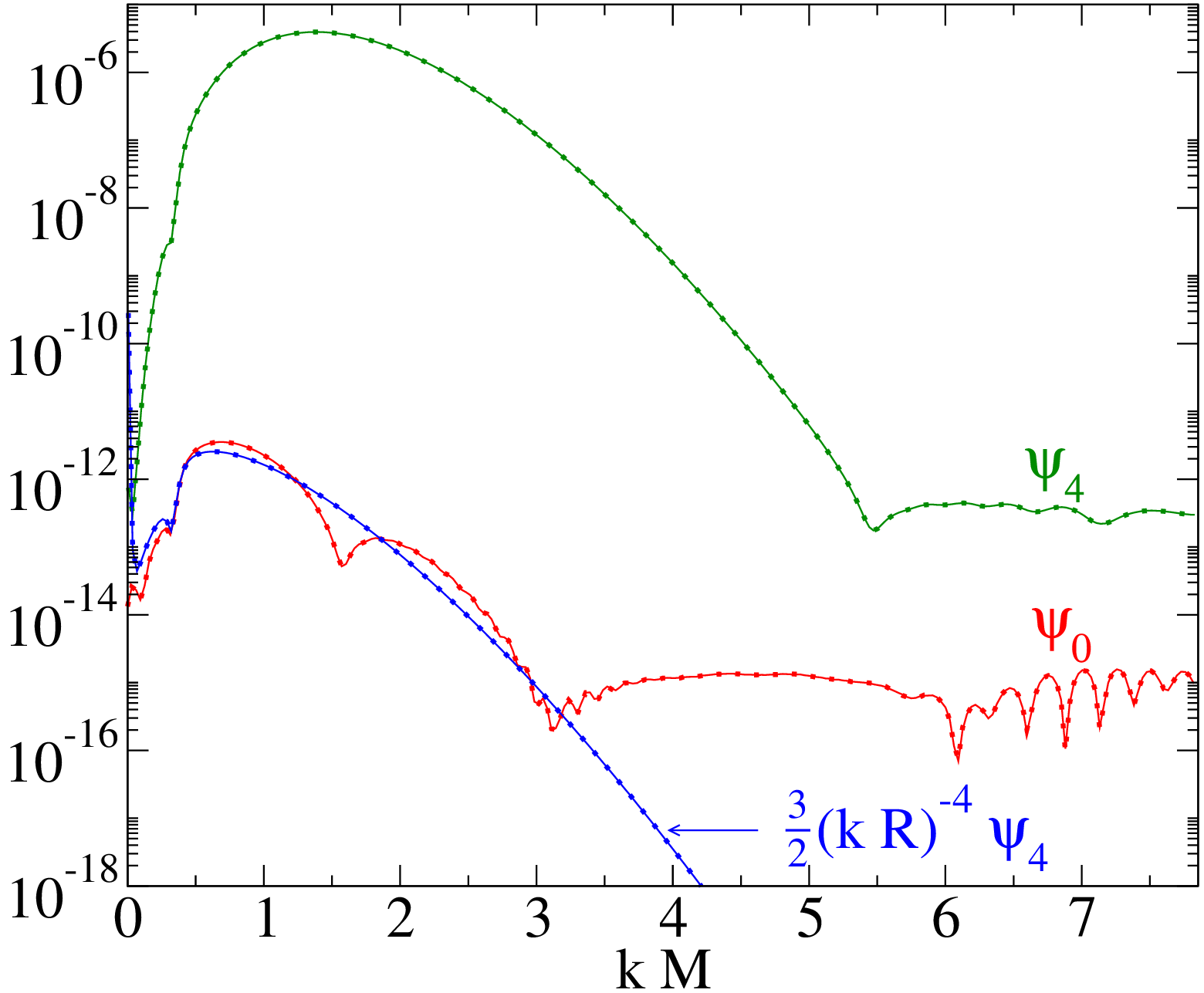}
  \plot{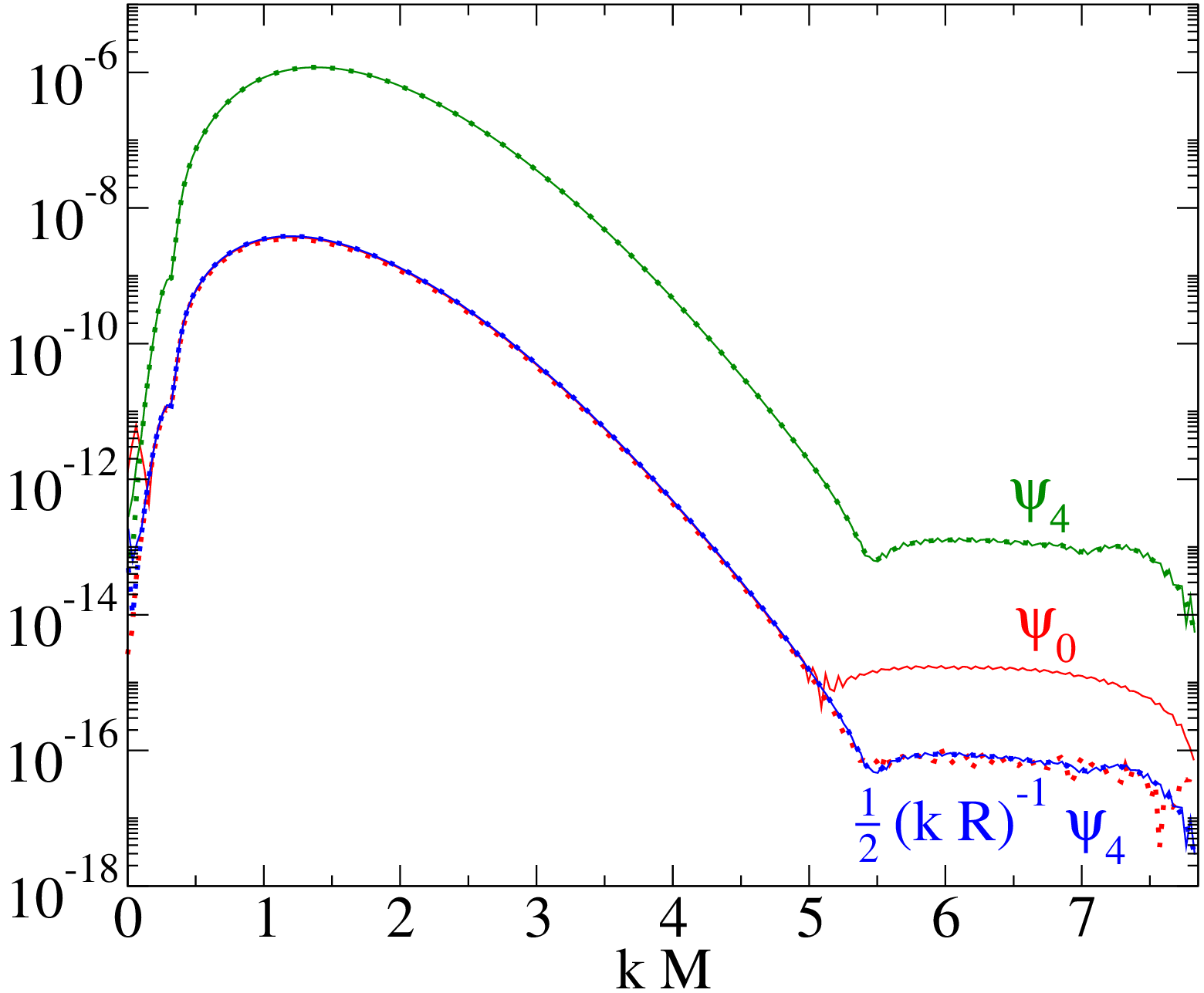}
  \plot{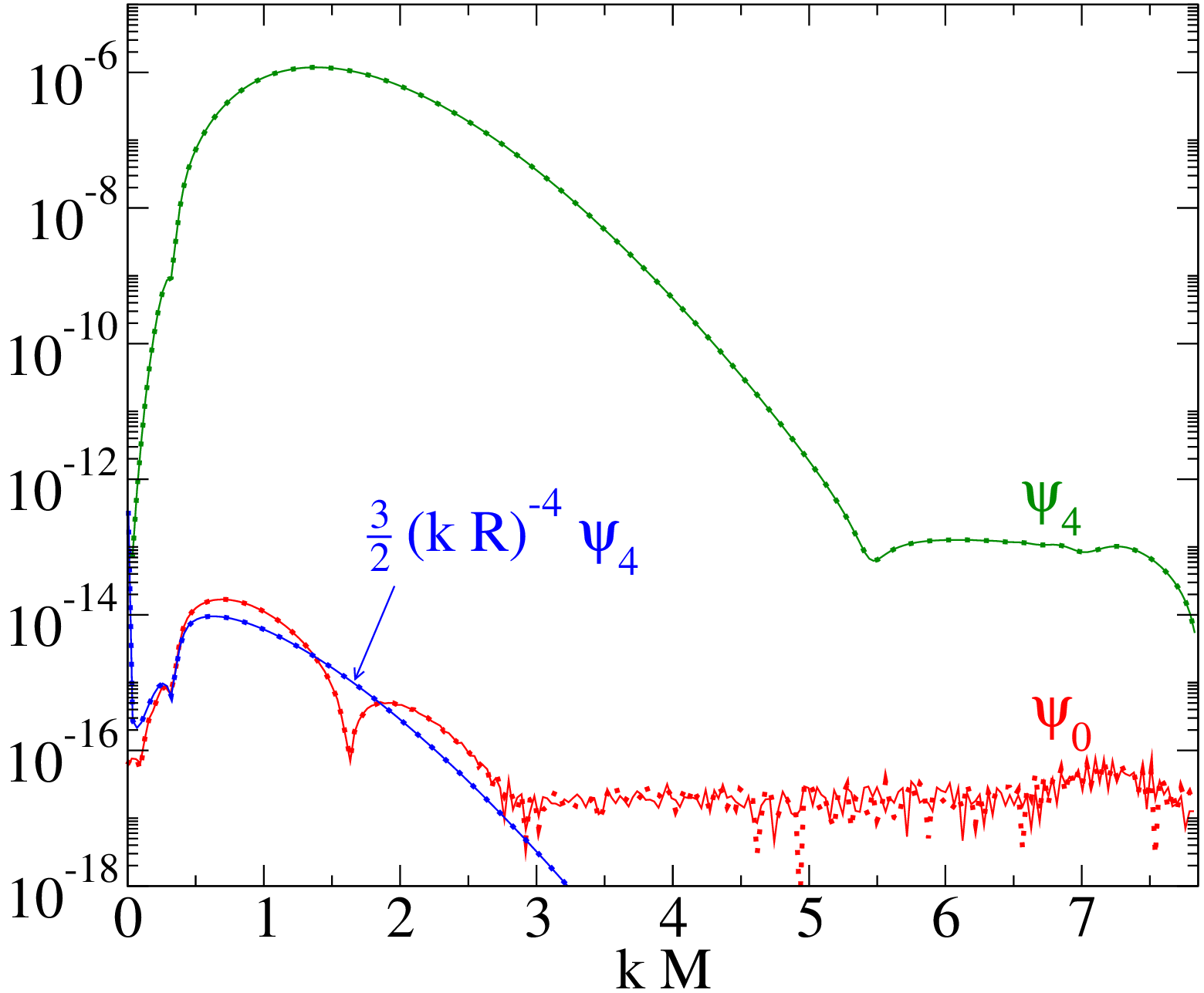}
  \caption{\label{fig:CompareReflCoeff} 
    Comparison of the time Fourier transform of the measured $\Psi_0(t)$ 
    (extracted $1.9 M$ in from the outer boundary)
    with the predicted value using the reflection coefficients derived
    in Sec.~\ref{Sect:ReflCoeffShearBC} and in \cite{Buchman:2006xf}.
    Left: 1.~vanishing shear (solid) vs.~2.~Kreiss-Winicour (dotted) 
    boundary conditions,
    right: 3.~second order boundary conditions (solid) vs.~4.~second order
    boundary conditions with first order gauge boundary conditions (dotted).
    Top: outer boundary radius $R = 41.9 M$, bottom: $R = 121.9 M$.
}
\end{figure}  

Finally we estimate the reflection coefficients for the various
boundary conditions numerically and compare with the analytical
predictions.  As a consequence of the results of Ref.~\cite{Buchman:2006xf},
the reflection coefficient can be approximated by forming the ratio of
the Newman-Penrose scalars $\Psi_0$ and $\Psi_4$ at the outer boundary,
\begin{equation}
  \label{Eq:ReflCoeffFromNP}
  |\gamma(kR)| = \frac{|\Psi_0|}{|\Psi_4|} + \Or(kR)^{-1}, 
\end{equation}
where $k$ is the wavenumber and $R$ is the boundary radius.
For the vanishing shear boundary conditions (\ref{Eq:FirstOrderBCList4}), 
we found in Sec.~\ref{Sect:ReflCoeffShearBC}
\begin{equation}
  |\gamma(kR)| = \half (kR)^{-1} + \Or(kR)^{-2},
\end{equation}
whereas for the freezing-$\Psi_0$ condition (\ref{Eq:SecondOrderBCList4}),
we have the much smaller reflection coefficient \cite{Buchman:2006xf}
\begin{equation}
  |\gamma(kR)| = \textstyle \frac{3}{2} (kR)^{-4} + \Or(kR)^{-5}.
\end{equation}
In Fig.~\ref{fig:CompareReflCoeff}, we compare the measured $\Psi_0$
with the predicted value obtained from Eq.~(\ref{Eq:ReflCoeffFromNP}),
using the measured $\Psi_4$ and the above analytical expressions for
the reflection coefficients.  A Fourier transform in time has been taken
in order to obtain plots vs.~wavenumber $k$.  The agreement is rather
good, roughly at the expected level of accuracy $\Or(kR)^{-1}$.
The leveling off of the numerical $\Psi_0$ for large $k$ 
is likely to be caused by numerical roundoff error 
(note the magnitude of $\Psi_0$ at large $k$).
The plots also indicate that the reflection coefficients are virtually
the same for both versions of the second-order boundary conditions 
(as expected since they only differ in the gauge boundary conditions),
and that the reflection coefficient of the original Kreiss-Winicour
boundary conditions agrees with that of our vanishing shear conditions.

Summarizing, both versions of the first order conditions (those
including the vanishing shear condition (\ref{Eq:FirstOrderBCList4})
and the original Kreiss-Winicour conditions) performed very similarly
in our numerical test. In contrast, the second order conditions caused
substantially less spurious reflections from the outer boundary.

%%%%%%%%%%%%%%%%%%%%%%%%%%%%%%%%%%%%%%%%%%%%%%%%%%%%%%%%%%%%%%
\section{Conclusions}
\label{Sect:Conclusions}
%%%%%%%%%%%%%%%%%%%%%%%%%%%%%%%%%%%%%%%%%%%%%%%%%%%%%%%%%%%%%%

In this paper, we have derived various sets of absorbing and
constraint-preserving boundary conditions for the Einstein equations
in the generalized harmonic gauge. We divided them into first, second
and higher order boundary conditions where the order refers to the
highest number of derivatives of the metric fields appearing in the
boundary conditions. The first order boundary conditions are a
generalization of the conditions considered by Kreiss and Winicour
\cite{Kreiss:2006mi} and specify the shear of the outgoing null
congruence associated with the two-dimensional cross sections of the
boundary surface. Our second order conditions enable one to fix the
Weyl scalar $\Psi_0$ at the boundary. Although there is a gauge
ambiguity in the definition of $\Psi_0$ at finite radius, these
conditions allow, in some sense, control of the incoming gravitational
radiation. This is important for simulations aimed at the far-field
extraction of gravitational waves emitted from compact astrophysical
sources. Furthermore, we could for example study the critical collapse
of gravitational waves by starting with Minkowski spacetime and
injecting pulses of gravitational radiation through the outer boundary
with different amplitudes \cite{Sarbach:2004rv,Lindblom06}. Finally,
we have considered higher order boundary conditions which comprise the
hierarchy of absorbing boundary conditions ${\cal B}_L$ and ${\cal
C}_L$ discussed in \cite{Buchman:2006xf,Buchman:2007pj}. As was shown
in these references, ${\cal B}_L$ and ${\cal C}_L$ yield fewer and
fewer spurious reflections of gravitational radiation as $L$ is
increased.

In Sec.~\ref{Sect:WellPosedness}, we have analyzed the well posedness
of the IBVPs resulting from our different boundary conditions. In
order to do so, we considered high-frequency perturbations of a given
smooth background solution in which case the problem reduces to a
system of ten decoupled wave equations with boundary conditions on a
frozen background spacetime. By means of a suitable coordinate
transformation, we have reduced the background metric to the flat
metric, with the exception of the component of the shift normal to the
boundary. Using the technique of Kreiss and Winicour
\cite{Kreiss:2006mi} which is based on a reduction to a
pseudo-differential first order system and the construction of smooth
symmetrizer, we then have shown that the resulting IBVPs are well
posed in the high-frequency limit. In view of the theory of
pseudo-differential operators \cite{Taylor99b} and the fact that we
obtain estimates for derivatives of arbirtrary order it is expected
that the full nonlinear problem is well posed as well. Our results
thus generalize the work of Ref.~\cite{Kreiss:2006mi} to non-trivial
shifts and boundary conditions of arbitrarily high order. They also
strengthen the result of Ref.~\cite{Rinne:2006vv}, where boundary
stability but not well posedness was proved for a first order version
of the generalized harmonic Einstein equations derived in
\cite{Lindblom06}. We remark that our results imply well posedness of
such first order formulations provided that the evolution system of
the additional constraints related to the first order reduction
(supplemented with suitable constraint-preserving boundary conditions)
is well posed. For a recent proof of well posedness for the first
order boundary conditions which is based on integration by parts, and
which does not require the pseudo-differential calculus, see
\cite{Kreiss:2007cc}.

In order to study the quality of the different boundary conditions
considered in this paper, we have computed the amount of spurious
gravitational radiation reflected off the boundary in the
high-frequency approximation in Sec.~\ref{Sect:Quality}. We have shown
that fewer and fewer reflections are present if the order of the
boundary conditions is increased. In addition, we have generalized
that analysis without the high frequency approximation for odd-parity
linear gravitational waves with wavenumber $k$ propagating on the
asymptotic region of a Schwarzschild background. For the case of a
spherical outer boundary of areal radius $R$ with the shear boundary
condition the reflection coefficient has been found to scale only as
$(kR)^{-1}$ for large $k R$ which is much slower than the $(kR)^{-4}$
decay calculated for the freezing-$\Psi_0$ boundary condition
\cite{Buchman:2006xf}. Finally, we have performed numerical tests of
some of our boundary conditions similar to the ones presented in
\cite{Rinne:2007ui}. The initial data were taken to be a Schwarzschild
black hole with an outgoing odd-parity quadrupolar gravitational wave
perturbation. The first order boundary conditions (with our modified
vanishing-shear condition) performed very similarly to the original
conditions considered in \cite{Kreiss:2006mi}. In contrast, as
expected from the analytic considerations, the second order conditions
caused substantially less spurious reflections from the outer
boundary. A numerical implementation of the higher order boundary
conditions is beyond the scope of this article and will be presented
in future work.

%%%%%%%%%%%%%%%%%%%%%%%%%%%
%%%   ACKNOWLEDGMENTS   %%%
%%%%%%%%%%%%%%%%%%%%%%%%%%%

\acknowledgments

It is a pleasure to thank J. Bardeen, L. Buchman, L. Lindblom,
O. Reula, M. Scheel and J. Winicour for useful comments and
discussions. The numerical simulations presented here were performed
using the Spectral Einstein Code (SpEC) developed at Caltech and
Cornell primarily by Larry Kidder, Mark Scheel and Harald
Pfeiffer. This work was supported in part by Direcci\'on General de
Estudios de Posgrado (DEGP), by CONACyT through grants 47201-F and
CONACYT 47209-F, by DGAPA-UNAM through grants IN113907, by grants CIC
4.20 to Universidad Michoacana, and by grants to Caltech from the
Sherman Fairchild Foundation, NSF grant PHY-0601459, and NASA grant
NNG05GG52G. M. Ruiz thanks Universidad Michoacana de San Nicol\a'as de
Hidalgo for hospitality.

%%%%%%%%%%%%%%%%%%%%%%
%%%   REFERENCES   %%%
%%%%%%%%%%%%%%%%%%%%%%

%\bibliographystyle{bibtex/apsrev}
\bibliography{bibtex/referencias}

\begin{thebibliography}{10}

\bibitem{Givoli91}
D.~Givoli.
\newblock Non-reflecting boundary conditions.
\newblock {\em J. Comp. Phys.}, 94:1--29, 1991.

\bibitem{Szilagyi:2002kv}
Bela Szilagyi and Jeffrey Winicour.
\newblock Well-posed initial-boundary evolution in general relativity.
\newblock {\em Phys. Rev.}, D68:041501, 2003.

\bibitem{Szilagyi:2001fy}
Bela Szilagyi, Bernd~G. Schmidt, and Jeffrey Winicour.
\newblock Boundary conditions in linearized harmonic gravity.
\newblock {\em Phys. Rev.}, D65:064015, 2002.

\bibitem{Lindblom06}
L.~Lindblom, M.~A. Scheel, L.~E. Kidder, R.~Owen, and O.~Rinne.
\newblock A new generalized harmonic evolution system.
\newblock {\em Class. Quant. Grav.}, 23:S447-- S462, 2006.

\bibitem{Kreiss:2006mi}
H.~O. Kreiss and J.~Winicour.
\newblock Problems which are well-posed in a generalized sense with
  applications to the {E}instein equations.
\newblock {\em Class. Quant. Grav.}, 23:S405--S420, 2006.

\bibitem{Babiuc:2006wk}
Maria~C. Babiuc, Bela Szilagyi, and Jeffrey Winicour.
\newblock Harmonic initial-boundary evolution in general relativity.
\newblock {\em Phys. Rev.}, D73:064017, 2006.

\bibitem{Motamed:2006uw}
Mohammad Motamed, M.~Babiuc, B.~Szilagyi, and H-O. Kreiss.
\newblock Finite difference schemes for second order systems describing black
  holes.
\newblock {\em Phys. Rev.}, D73:124008, 2006.

\bibitem{Babiuc:2006ik}
M.~C. Babiuc, H.~O. Kreiss, and Jeffrey Winicour.
\newblock Constraint-preserving {S}ommerfeld conditions for the harmonic
  {E}instein equations.
\newblock {\em Phys. Rev.}, D75:044002, 2007.

\bibitem{Rinne:2006vv}
Oliver Rinne.
\newblock Stable radiation-controlling boundary conditions for the generalized
  harmonic {E}instein equations.
\newblock {\em Class. Quant. Grav.}, 23:6275--6300, 2006.

\bibitem{Rinne:2007ui}
Oliver Rinne, Lee Lindblom, and Mark~A. Scheel.
\newblock Testing outer boundary treatments for the {E}instein equations.
\newblock {\em Class. Quant. Grav.}, 24:4053--4078, 2007.

\bibitem{Buchman:2006xf}
Luisa~T. Buchman and Olivier C.~A. Sarbach.
\newblock Towards absorbing outer boundaries in general relativity.
\newblock {\em Class. Quant. Grav.}, 23:6709--6744, 2006.

\bibitem{Buchman:2007pj}
Luisa~T. Buchman and Olivier C.~A. Sarbach.
\newblock Improved outer boundary conditions for {E}instein's field equations.
\newblock {\em Class. Quant. Grav.}, 24:S307--S326, 2007.

\bibitem{Friedrich99}
H.~Friedrich and G.~Nagy.
\newblock The initial boundary value problem for {E}instein's vacuum field
  equations.
\newblock {\em Comm. Math. Phys.}, 201:619--655, 1999.

\bibitem{Friedrichs58}
K.O. Friedrichs.
\newblock Symmetric positive linear differential equations.
\newblock {\em Commun. Pure Appl. Math.}, 11:333--418, 1958.

\bibitem{Lax60}
P.D. Lax and R.S. Phillips.
\newblock Local boundary conditions for dissipative symmetric linear
  differential operators.
\newblock {\em Commun. Pure Appl. Math.}, 13:427--455, 1960.

\bibitem{Secchi96}
P.~Secchi.
\newblock Well-posedness of characteristic symmetric hyperbolic systems.
\newblock {\em Arch. Rat. Mech. Anal.}, 134:155--197, 1996.

\bibitem{Calabrese:2002xy}
Gioel Calabrese, Jorge Pullin, Olivier Sarbach, Manuel Tiglio, and Oscar Reula.
\newblock Well posed constraint-preserving boundary conditions for the
  linearized {E}instein equations.
\newblock {\em Commun. Math. Phys.}, 240:377--395, 2003.

\bibitem{Gundlach:2004jp}
Carsten Gundlach and Jose~M. Mart\a'{\i}n-Garc\a'{\i}a.
\newblock Symmetric hyperbolicity and consistent boundary conditions for
  second-order {E}instein equations.
\newblock {\em Phys. Rev.}, D70:044032, 2004.

\bibitem{Nagy:2006pr}
Gabriel Nagy and Olivier Sarbach.
\newblock A minimization problem for the lapse and the initial- boundary value
  problem for {E}instein's field equations.
\newblock {\em Class. Quant. Grav.}, 23:S477--S504, 2006.

\bibitem{Kreiss89}
H.~O. Kreiss and J.~Lorenz.
\newblock {\em Initial-boundary value problems and the {N}avier-{S}tokes
  equations}.
\newblock Academic Press, San Diego, 1989.

\bibitem{Kreiss70}
H.O. Kreiss.
\newblock Initial boundary value problems for hyperbolic systems.
\newblock {\em Commun. Pure Appl. Math.}, 23:277--298, 1970.

\bibitem{Stewart98}
J.M. Stewart.
\newblock The {C}auchy problem and the initial boundary value problem in
  numerical relativity.
\newblock {\em Class. Quantum Grav.}, 15:2865--2889, 1998.

\bibitem{Majda75}
A.~Majda and S.~Osher.
\newblock Initial-boundary value problems for hyperbolic equations with
  uniformly characteristic boundary.
\newblock {\em Commun.\ Pure Appl.\ Math.}, 28:607--675, 1975.

\bibitem{Taylor99b}
M.E. Taylor.
\newblock {\em Partial differential equations II, Qualitative Studies of Linear
  Equations}.
\newblock Springer, 1999.

\bibitem{Bardeen:2001xx}
James~M. Bardeen and L.~T. Buchman.
\newblock Numerical tests of evolution systems, gauge conditions, and boundary
  conditions for 1d colliding gravitational plane waves.
\newblock {\em Phys. Rev.}, D65:064037, 2002.

\bibitem{Sarbach:2004rv}
Olivier Sarbach and Manuel Tiglio.
\newblock Boundary conditions for {E}instein's field equations: Analytical and
  numerical analysis.
\newblock {\em J. Hyperbol. Diff. Equat.}, 2:839, 2005.

\bibitem{Lindblom05}
L.E. Kidder, L.~Lindblom, M.A. Scheel, L.T. Buchman, and H.P. Pfeiffer.
\newblock Boundary conditions for the {E}instein evolution system.
\newblock {\em Phys. Rev. D}, 71:064020, 2005.

\bibitem{Friedrich85}
H.~Friedrich.
\newblock On the hyperbolicity of {E}instein's and other gauge field equations.
\newblock {\em Comm. Math. Phys.}, 100:525--543, 1985.

\bibitem{Friedrich96}
H.~Friedrich.
\newblock Hyperbolic reductions for {E}instein's equations.
\newblock {\em Class. Quantum Grav.}, 13:1451--1469, 1996.

\bibitem{Andersson03}
L.~Andersson and V.~Moncrief.
\newblock Elliptic-hyperbolic systems and the {E}instein equations.
\newblock {\em Annales Henri Poincar\a'e}, 4:1--34, 2003.

\bibitem{Pretorius:2004jg}
Frans Pretorius.
\newblock Numerical relativity using a generalized harmonic decomposition.
\newblock {\em Class. Quant. Grav.}, 22:425--452, 2005.

\bibitem{Fischer72}
A.~E. Fischer and J.~E. Marsden.
\newblock The {E}instein evolution equations as a first-order quasi-linear
  symmetric hyperbolic system, {I}.
\newblock {\em Comm. Math. Phys.}, 28:1--38, 1972.

\bibitem{Foures-Bruhat52}
Y.~Foures-Bruhat.
\newblock Th\'eor\`eme d'{\'e}xistence pour certains syst\`emes d'\'equations
  aux d\'eriv\'ees partielles non lin\'eaires.
\newblock {\em Acta Math.}, 88:141--225, 1952.

\bibitem{Gustafsson95}
B.~Gustafsson, H.O. Kreiss, and J.~Oliger.
\newblock {\em Time dependent problems and difference methods}.
\newblock Wiley, New York, 1995.

\bibitem{Reula:2004nr}
Oscar Reula and Olivier Sarbach.
\newblock A model problem for the initial-boundary value formulation of
  {E}instein's field equations.
\newblock {\em J. Hyperbol. Diff. Equat.}, 2:397--435, 2005.

\bibitem{Engquist77}
B.~Engquist and A.~Majda.
\newblock Absorbing boundary conditions for the numerical simulation of waves.
\newblock {\em Math. Comp.}, 31:629--651, 1977.

\bibitem{Gerlach79}
U.H. Gerlach and U.K. Sengupta.
\newblock Gauge-invariant perturbations on most general spherically symmetric
  space-times.
\newblock {\em Phys. Rev. D}, 19:2268--2272, 1979.

\bibitem{Regge57}
T.~Regge and J.~Wheeler.
\newblock Stability of a {S}chwarzschild singularity.
\newblock {\em Phys. Rev.}, 108:1063--1069, 1957.

\bibitem{Sarbach:1999gi}
Olivier Sarbach, Markus Heusler, and Othmar Brodbeck.
\newblock Perturbation theory for self-gravitating gauge fields: The odd-parity
  sector.
\newblock {\em Phys. Rev.}, D62:084001, 2000.

\bibitem{Sarbach:2001qq}
Olivier Sarbach and Manuel Tiglio.
\newblock Gauge invariant perturbations of {S}chwarzschild black holes in
  horizon-penetrating coordinates.
\newblock {\em Phys. Rev.}, D64:084016, 2001.

\bibitem{Kreiss:2007cc}
H.~O. Kreiss, O.~Reula, O.~Sarbach, and J.~Winicour.
\newblock Well-posed initial-boundary value problem for the harmonic {E}instein
  equations using energy estimates.
\newblock {\em Class. Quant. Grav.}, 2007.
\newblock To appear.

\end{thebibliography}

%%%%%%%%%%%%%%%
%%%   END   %%%
%%%%%%%%%%%%%%%

\end{document}